\documentclass[twocolumn,aps,pra]{revtex4-2}
\usepackage{bm}    % bold math
\usepackage{amssymb,amsmath}
\usepackage{times}
\usepackage{graphicx}
\usepackage[colorlinks=true,linkcolor=blue,citecolor=blue,urlcolor=blue]{hyperref}

\begin{document}

\title{Thresholdless dynamical instability in transverse $\mathcal{PT}$-symmetric scattering systems: \\ The hidden role of bound states in the continuum}

\author{Chao Zheng}
\email{zhengchaonju@gmail.com}
\affiliation{School of Education, Jiangsu Open University, Nanjing 210036, China}

\begin{abstract}

The stationary scattering properties of transverse parity-time ($\mathcal{PT}$) symmetric systems have been extensively studied, yet their dynamical stability, a prerequisite for any stationary description, remains largely unexplored.
Here we uncover a thresholdless dynamical instability in such systems, driven by symmetry-protected bound states in the continuum (BICs).
Without gain and loss, the up-down mirror symmetry of the transverse geometry generically protects a BIC, which manifests as the coalescence of an $S$-matrix pole and zero on the real axis of the complex wave-number plane.
A Hermitian symmetry-breaking perturbation shifts the poles into the lower half-plane, converting the BIC into a resonance with a Fermi-golden-rule decay width.
An anti-Hermitian $\mathcal{PT}$-symmetric perturbation instead reverses the sign of the second-order energy shift, driving the poles into the upper half-plane and producing a time-growing bound state with $\operatorname{Im}E=\gamma^{2}\Gamma_{V}/2+O(\gamma^{3})$, where $\gamma$ is the gain-loss strength and $\Gamma_{V}$ is the golden-rule coupling of the BIC to the continuum.
The instability therefore sets in at arbitrarily small $\gamma$ whenever $\Gamma_{V}>0$.
We confirm this mechanism in a three-site side-coupled model that is unstable despite possessing a unitary scattering matrix, and in a four-site rhombic model where transverse and longitudinal gain-loss placements yield vanishing and finite thresholds, respectively.
These results establish a microscopic stability criterion and a design principle for stable $\mathcal{PT}$-symmetric scattering devices.

\end{abstract}

\date{\today}

\maketitle

%%%%%%%%%%%%%%%%%%%%%%%%%%%%%%%%%%%%%%%%%%%%%%%%%%%%%%%%%%%%%%%%%%%%%%%%%%%%%%%%%%%%%%%%%%%%%%%%%%%%
\section{Introduction}
\label{sec:Introduction}
%%%%%%%%%%%%%%%%%%%%%%%%%%%%%%%%%%%%%%%%%%%%%%%%%%%%%%%%%%%%%%%%%%%%%%%%%%%%%%%%%%%%%%%%%%%%%%%%%%%%

Parity-time ($\mathcal{PT}$) symmetric systems are non-Hermitian yet invariant under the combined action of spatial reflection and time reversal.
Their energy spectra can remain entirely real below a critical non-Hermiticity threshold, beyond which $\mathcal{PT}$ symmetry is spontaneously broken and eigenvalues form complex-conjugate pairs~\cite{bender1998a,bender2024}.
This remarkable property has stimulated extensive theoretical and experimental work across optics, photonics, acoustics, and electronic circuits, revealing wave phenomena and device functionalities that have no counterpart in Hermitian systems~\cite{konotop2016,feng2017,el-ganainy2018,ozdemir2019}.

\begin{figure}
\centering
\includegraphics{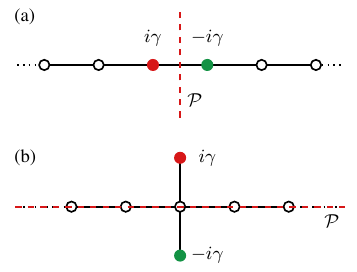}
\caption{
Schematic illustration of two classes of $\mathcal{PT}$-symmetric scattering systems, distinguished by the orientation of the parity mirror axis $\mathcal{P}$ (red dashed line) relative to the transport direction.
(a) Longitudinal configuration, where the mirror axis is perpendicular to the transport direction, so that the balanced gain $+i\gamma$ (red) and loss $-i\gamma$ (green) are distributed along the propagation axis.
(b) Transverse configuration, where the mirror axis coincides with the transport direction, so that the balanced gain and loss reside on opposite sides of the propagation channel.
}
\label{fig:schematic}
\end{figure}

In scattering settings, the orientation of the parity mirror plane relative to the transport direction distinguishes two classes of $\mathcal{PT}$-symmetric systems, as illustrated in Fig.~\ref{fig:schematic}~\cite{jin2016a}. 
In a longitudinal configuration, the mirror plane is perpendicular to the transport direction, so that gain and loss are distributed along the propagation axis [Fig.~\ref{fig:schematic}(a)].
Left- and right-incident waves traverse the gain-loss profile in reverse order, rendering reflection generically nonreciprocal while transmission remains reciprocal~\cite{ramezani2014,garmon2015,jin2016a,zhu2016,achilleos2017a,shobe2021}.
The scattering amplitudes obey a generalized unitarity relation~\cite{ge2012}, enabling phenomena such as unidirectional invisibility~\cite{lin2011,feng2013} and simultaneous coherent perfect absorption and lasing~\cite{longhi2010,chong2011}.
In a transverse configuration, the mirror plane contains the transport direction, placing balanced gain and loss on opposite sides of the propagation channel [Fig.~\ref{fig:schematic}(b)].
A wave traveling in either direction experiences the same gain-loss profile, making reflection reciprocal.
Moreover, the scattering matrix ($S$ matrix) can remain unitary despite the non-Hermiticity of the scattering center~\cite{jin2012,zhu2015,jin2016a,jin2022,ganguly2022,soori2023}.
This geometry arises naturally in systems with side-coupled defects~\cite{miroshnichenko2011a,jin2012,zhu2015,zhang2019c,zhang2019d,li2020a,niu2026}, parallel quantum dots~\cite{jin2016a,li2017,zhang2017a,zhang2017b,zhang2020e,soori2023}, quantum rings~\cite{ganguly2022}, and ladder structures~\cite{soori2023}, where the interplay of gain, loss, and quantum interference can convert perfect reflection into perfect transmission~\cite{miroshnichenko2011a,zhu2015,zhang2017b}, shift Fano resonances~\cite{zhang2017b,zhang2019d,li2020a}, or create and annihilate resonances~\cite{zhang2017a,zhang2017b,zhang2019c,zhang2019d,zhang2020e,ganguly2022,niu2026}.

Despite this rich phenomenology, existing studies of $\mathcal{PT}$-symmetric scattering systems have focused almost exclusively on stationary transport properties, while their dynamical stability has been largely overlooked.
Stationary scattering coefficients are determined by continuum scattering states at real energies, whereas dynamical stability is governed by discrete bound states, which can lie at complex energies in non-Hermitian systems.
If any bound state acquires an energy with a positive imaginary part, it grows exponentially in time, eventually overwhelming the scattering response and rendering the stationary description unphysical~\cite{zheng2025a}. 
Dynamical stability is therefore a prerequisite for any $\mathcal{PT}$-symmetric scattering device to function as designed. 
For longitudinal configurations, this issue has been examined in Refs.~\cite{bendix2009,dmitriev2011,zheng2026}; for the transverse case, however, no systematic investigation has been carried out.

In this work, we uncover a thresholdless dynamical instability of transverse $\mathcal{PT}$-symmetric scattering systems, driven by symmetry-protected bound states in the continuum (BICs) of the underlying Hermitian structure.
In the absence of gain and loss, the up-down mirror symmetry of the transverse geometry generically protects a BIC that is completely decoupled from the scattering continuum.
In the $S$-matrix formalism, this BIC appears as the exact coalescence of a pole and a zero on the real axis of the complex $k$ plane.
A Hermitian perturbation that breaks the mirror symmetry couples the BIC to the continuum and shifts the poles into the lower half-plane, yielding a resonant and an antiresonant state with a width given by Fermi's golden rule.
In contrast, an anti-Hermitian $\mathcal{PT}$-symmetric perturbation reverses the sign of the second-order energy shift and drives the poles into the upper half-plane, yielding a time-growing and a time-decaying bound state.
The instability therefore sets in at arbitrarily small gain-loss strength whenever the golden-rule coupling between the BIC and the continuum is nonzero.
We confirm this mechanism through exact pole trajectories and wave-packet dynamics in two tight-binding models.
A three-site side-coupled model exhibits the thresholdless instability even though its $S$ matrix remains unitary at all gain-loss strengths.
A four-site rhombic model shows that transverse and longitudinal placements of the gain-loss potential yield a vanishing and a finite instability threshold, respectively.
These results establish a microscopic stability criterion and identify the spatial overlap between the gain-loss potential and the BIC wave function as a key design parameter for stable $\mathcal{PT}$-symmetric scattering devices.

The remainder of this paper is organized as follows.
Section~\ref{sec:formalism} presents the general formalism and derives the thresholdless instability criterion.
Section~\ref{sec:examples} illustrates the mechanism with the three-site side-coupled model and the four-site rhombic model.
Section~\ref{sec:conclusions} summarizes and discusses our findings.

%%%%%%%%%%%%%%%%%%%%%%%%%%%%%%%%%%%%%%%%%%%%%%%%%%%%%%%%%%%%%%%%%%%%%%%%%%%%%%%%%%%%%%%%%%%%%%%%%%%%
\section{General formalism}
\label{sec:formalism}
%%%%%%%%%%%%%%%%%%%%%%%%%%%%%%%%%%%%%%%%%%%%%%%%%%%%%%%%%%%%%%%%%%%%%%%%%%%%%%%%%%%%%%%%%%%%%%%%%%%%

%%%%%%%%%%%%%%%%%%%%%%%%%%%%%%%%%%%%%%%%%%%%%%%%%%%%%%%%%%%%%%%%%%%%%%%%%%%%%%%%%%%%%%%%%%%%%%%%%%%%
\subsection{Model and symmetry-protected BIC}
%%%%%%%%%%%%%%%%%%%%%%%%%%%%%%%%%%%%%%%%%%%%%%%%%%%%%%%%%%%%%%%%%%%%%%%%%%%%%%%%%%%%%%%%%%%%%%%%%%%%

We consider a Hermitian scattering center with $N$ internal sites coupled to two semi-infinite tight-binding leads,
\begin{equation}
H_0=H_L+H_R+H_c+H_{\mathrm{int}}.
\label{eq:H0}
\end{equation}
Here $H_{L,R}=-J\sum_{j}(|j\rangle\langle j+1|+\mathrm{H.c.})$ denote the left and right leads with hopping amplitude $J$, $H_{\mathrm{c}}$ describes the scattering center, and $H_{\mathrm{int}}$ couples the center to the terminal lead sites.
The leads support plane-wave solutions with dispersion relation
\begin{equation}
	E(k)=-2J\cos k
\label{eq:dispersion}
\end{equation}
for $k\in[-\pi,\pi]$, defining a scattering continuum $|E|\le 2J$.

\begin{figure}
\centering
\includegraphics{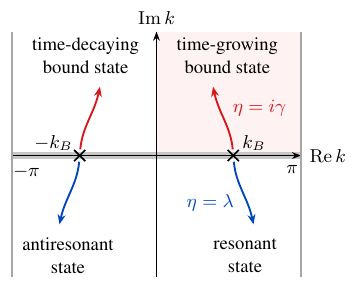}
\caption{
Trajectories of the $S$-matrix poles associated with a BIC in the complex-$k$ plane under mirror-symmetry-breaking perturbations.
Poles in each quadrant correspond to physically distinct states as labeled. 
The gray band marks the scattering continuum.
The unperturbed BIC appears as pole-zero coalescences on the real axis at $k=\pm k_B$ (black crosses).
A Hermitian perturbation (blue arrows) shifts the poles into the lower half-plane, producing a resonant and an antiresonant state. 
A $\mathcal{PT}$-symmetric perturbation (red arrows) drives the poles into the upper half-plane, producing a time-growing and a time-decaying bound state.
The time-growing bound state in the first quadrant (shaded) renders the system dynamically unstable.
}
\label{fig:pole_trajectory}
\end{figure}

We require $H_0$ to possess an up-down mirror symmetry $\mathcal{P}$ satisfying $[\mathcal{P},H_0]=0$.
This symmetry is naturally present in transverse $\mathcal{PT}$-symmetric scattering systems before the gain-loss potential is introduced [Fig.~\ref{fig:schematic}(b)].
Since both leads lie on the mirror axis, $\mathcal{P}$ acts as the identity on all lead sites, so every extended scattering state is necessarily parity-even.
A parity-odd eigenstate, by contrast, must have vanishing amplitude on all lead sites and is therefore confined entirely within the scattering center.
Suppose the center supports such an eigenstate $|B\rangle$ at an energy $E_B$ within the continuum,
\begin{equation}
    H_0|B\rangle = E_B|B\rangle, \quad \mathcal{P}|B\rangle = -|B\rangle,
\end{equation}
the parity mismatch forbids any coupling between $|B\rangle$ and the propagating modes.
The state $|B\rangle$ therefore remains perfectly localized despite being spectrally embedded in the continuum: it is a symmetry-protected BIC~\cite{ladrondeguevara2006,plotnik2011,hsu2016,sadreev2021a}.
Decoupled from all scattering channels, it carries zero radiative width and leaves no signature in the transmission or reflection spectrum.
For simplicity, we assume a single nondegenerate BIC; the generalization to multiple or degenerate BICs is straightforward.

In the $S$-matrix formalism, discrete states of the open system correspond to poles in the complex $k$ plane, obtained by solving the Schr\"{o}dinger equation with purely outgoing boundary conditions~\cite{siegert1939,hatano2013}.
A pole at $k_n = k_n^r + ik_n^i$ has energy $E_n = -2J\cos k_n \equiv E_n^r + iE_n^i$, with imaginary part $E_n^i = 2J\sin k_n^r\,\sinh k_n^i$.
The associated wave function behaves asymptotically as $e^{ik_n|x|}e^{-iE_nt}$ in the leads, so $k_n^i$ governs the spatial profile and $E_n^i$ governs the temporal evolution.
Poles in the upper half-plane ($k_n^i>0$) are spatially localized and normalizable, residing on the physical Riemann sheet of the complex energy surface~\cite{sasada2011,hatano2014,garmon2015}.
Among these, first-quadrant poles ($k_n^r>0$, hence $E_n^i>0$) represent time-growing bound states (TGBSs), while second-quadrant poles ($k_n^r<0$, hence $E_n^i<0$) represent time-decaying bound states.
Poles in the lower half-plane ($k_n^i<0$) are spatially divergent and reside on the unphysical Riemann sheet, with third-quadrant poles corresponding to antiresonant states and fourth-quadrant poles to resonant states.

A BIC at energy $E_B=E(k_B)$ manifests as the exact coalescence of an $S$-matrix pole and a zero on the real axis at $\pm k_B$ (see Fig.~\ref{fig:pole_trajectory})~\cite{krasnok2019}.
Breaking the mirror symmetry lifts this pole-zero cancellation and displaces the pole off the real axis.
When a pole enters the first quadrant, a TGBS emerges and the system becomes dynamically unstable~\cite{zheng2025a}.

%%%%%%%%%%%%%%%%%%%%%%%%%%%%%%%%%%%%%%%%%%%%%%%%%%%%%%%%%%%%%%%%%%%%%%%%%%%%%%%%%%%%%%%%%%%%%%%%%%%%
\subsection{Evolution of BIC under perturbation}
%%%%%%%%%%%%%%%%%%%%%%%%%%%%%%%%%%%%%%%%%%%%%%%%%%%%%%%%%%%%%%%%%%%%%%%%%%%%%%%%%%%%%%%%%%%%%%%%%%%%

We now break the mirror symmetry by adding a perturbation to the scattering center and track the resulting evolution of the $S$-matrix poles originating from the BIC.
The perturbed Hamiltonian reads
\begin{equation}
H = H_0 + \eta V,
\label{eq:H_perturbed}
\end{equation}
where $V$ is a Hermitian, parity-odd operator acting on the scattering center, satisfying $\mathcal{P}V\mathcal{P}^{-1} = -V$, and $\eta$ controls both the type and strength of the perturbation.
For real $\eta$, the perturbation is Hermitian.
For $\eta=i\gamma$ with real $\gamma$, the perturbation is anti-Hermitian and realizes a transverse $\mathcal{PT}$-symmetric gain-loss potential.

To obtain the pole energies emerging from the BIC,  we employ the Feshbach projection method.
Let $P=|B\rangle\langle B|$ be the projector onto the BIC subspace and $Q=I-P$ its orthogonal complement.
Projecting the eigenvalue equation $H|\Psi\rangle=E|\Psi\rangle$ onto these two subspaces and eliminating the $Q$ component yields an exact implicit equation for the pole energy~\cite{hubac2010},
\begin{equation}
\begin{aligned}
E ={} & E_B + \eta\,\langle B|V|B\rangle \\
      & + \eta^{2}\,\langle B|V\,Q\frac{1}{E-Q(H_0+\eta V)Q}\,Q\,V|B\rangle .
\end{aligned}
\label{eq:exact}
\end{equation}
Since both $V$ and $|B\rangle$ are parity-odd, the first-order correction vanishes,
\begin{equation}
\langle B|V|B\rangle
= \langle B|\mathcal{P}^{-1}(\mathcal{P}V\mathcal{P}^{-1})\mathcal{P}|B\rangle
= -\langle B|V|B\rangle = 0,
\label{eq:first_order}
\end{equation}
and the leading correction to the BIC energy is quadratic in $\eta$.
Expanding Eq.~\eqref{eq:exact} to second order by setting $E=E_B$ and dropping $\eta QVQ$ in the resolvent denominator, we obtain
\begin{equation}
E_\sigma = E_B + \eta^2\,\langle B|V\,R_\sigma(E_B)\,V|B\rangle + O(\eta^3),
\label{eq:second}
\end{equation}
where
\begin{equation}
R_\sigma(E_B) = Q\,\frac{1}{E_B - H_0 + i\sigma\,0^+}\,Q
\label{eq:Rsigma}
\end{equation}
is the projected resolvent of $H_0$.
The index $\sigma = \pm$ labels the two poles originating from $k = +k_B$ and $k = -k_B$, with the infinitesimal $i\sigma\,0^+$ implementing outgoing ($\sigma = +$) or incoming ($\sigma = -$) boundary conditions.

To evaluate the matrix element in Eq.~\eqref{eq:second}, we insert the resolution of identity for the $Q$ subspace,
\begin{equation}
	Q = \sum_\nu |\phi_\nu\rangle\langle\phi_\nu|
	+ \int_{-\pi}^{\pi} \frac{dk}{2\pi} \, |\psi_k\rangle\langle\psi_k| ,
	\label{eq:completeness}
\end{equation}
where $|\phi_\nu\rangle$ denotes the bound states of $H_0$ other than $|B\rangle$, whose energies $E_\nu$ lie outside the continuum, and $|\psi_k\rangle$ denotes the scattering states normalized as $\langle\psi_k|\psi_{k'}\rangle=2\pi\delta(k-k')$, with $k>0$ ($k<0$) labeling incidence from the left (right).
Applying the Sokhotski--Plemelj identity $(x+i\sigma\,0^+)^{-1}=\mathrm{P}\,x^{-1}-i\pi\sigma\,\delta(x)$ 
and using $\delta[E_B-E(k)]=v_B^{-1}\left[\delta(k-k_B)+\delta(k+k_B)\right]$
with $v_B=|dE/dk|_{\pm k_B}=2J\sin k_B>0$, we obtain
\begin{equation}
\langle B|V\,R_{\sigma}(E_B)\,V|B\rangle=\Delta_{V}-\frac{i}{2}\sigma\Gamma_{V},
\label{eq:DeltaC}
\end{equation}
with
\begin{equation}
\Delta_{V}
=\sum_{\nu}\frac{|\langle\phi_{\nu}|V|B\rangle|^{2}}{E_B-E_{\nu}}
+\mathrm{P}\!\!\int_{-\pi}^{\pi}\!\frac{dk}{2\pi}\,
\frac{|\langle\psi_{k}|V|B\rangle|^{2}}{E_B-E(k)}
\label{eq:Delta}
\end{equation}
and
\begin{equation}
\Gamma_{V}=\frac{1}{v_B}\sum_{\sigma=\pm}\bigl|\langle\psi_{\sigma k_B}|V|B\rangle\bigr|^{2}.
\label{eq:Gamma}
\end{equation}
Here $\Delta_V$ is the real level shift arising from virtual transitions to off-shell states, and $\Gamma_{V}$ is the Fermi-golden-rule coupling between the BIC and the propagating modes at energy $E_B$.

Substituting Eq.~\eqref{eq:DeltaC} into Eq.~\eqref{eq:second} gives the pole energies to leading order,
\begin{equation}
E_\sigma=E_B+\eta^2\Delta_V-\frac{i}{2}\sigma\eta^2\Gamma_V+O(\eta^3).
\label{eq:pole_energy}
\end{equation}
To map this result onto the complex $k$ plane, we linearize the dispersion $E = -2J\cos k$ near $k = \sigma k_B$, where the group velocity is $\sigma v_B$.
Inverting the linearized relation $E_\sigma-E_B\approx \sigma v_B(k_\sigma-\sigma k_B)$ yields
\begin{equation}
k_\sigma=\sigma k_B+\sigma\frac{\eta^2\Delta_V}{v_B}
-i\frac{\eta^2\Gamma_V}{2v_B}+O(\eta^3).
\label{eq:pole_k}
\end{equation}
The real part of $k_\sigma$ shifts in opposite directions for $\sigma=+$ and $\sigma=-$, while the imaginary part is independent of $\sigma$: both poles migrate vertically in the same direction.
Since $\Gamma_V\ge 0$, the sign of $\eta^2$ alone determines whether the poles descend into the lower half-plane or ascend into the upper half-plane, thereby fixing the physical character of the resulting states.

%%%%%%%%%%%%%%%%%%%%%%%%%%%%%%%%%%%%%%%%%%%%%%%%%%%%%%%%%%%%%%%%%%%%%%%%%%%%%%%%%%%%%%%%%%%%%%%%%%%%
\subsection{Thresholdless instability and its origin}
\label{subsec:instability}
%%%%%%%%%%%%%%%%%%%%%%%%%%%%%%%%%%%%%%%%%%%%%%%%%%%%%%%%%%%%%%%%%%%%%%%%%%%%%%%%%%%%%%%%%%%%%%%%%%%%

To highlight the distinctive role of the transverse $\mathcal{PT}$-symmetric potential, we first examine a Hermitian perturbation that breaks the mirror symmetry.
Setting $\eta=\lambda$ with $\lambda\in\mathbb{R}$ gives $\eta^2=\lambda^2>0$.
Equations~\eqref{eq:pole_energy} and \eqref{eq:pole_k} then yield
\begin{equation}
	\operatorname{Im} E_{\sigma} = -\frac{\sigma}{2}\lambda^{2}\Gamma_V+O(\lambda^3), \quad
	\operatorname{Im} k_{\sigma} = -\frac{\lambda^{2}\Gamma_V}{2 v_B}+O(\lambda^3).
	\label{eq:Im_Hermitian}
\end{equation}
For $\Gamma_V>0$, both poles descend into the lower half of the complex $k$ plane, producing a resonant state in the fourth quadrant and an antiresonant state in the third quadrant [Fig.~\ref{fig:pole_trajectory}].
This is the familiar fate of symmetry-protected BICs: breaking the protecting symmetry couples the dark state to the continuum and endows it with a radiative decay rate $\lambda^{2}\Gamma_V$ given by Fermi's golden rule.
The BIC is thereby converted into a quasi-bound state that manifests as a Fano resonance in the transmission spectrum~\cite{guevara2003,miroshnichenko2009a,miroshnichenko2010,sadreev2021a}.

The situation changes qualitatively for a transverse $\mathcal{PT}$-symmetric perturbation.
Setting $\eta=i\gamma$ with $\gamma\in\mathbb{R}$ yields $\eta^{2}=-\gamma^{2}<0$, which reverses the sign of the second-order correction in Eq.~\eqref{eq:second}.
Equations~\eqref{eq:pole_energy} and \eqref{eq:pole_k} now give
\begin{equation}
	\operatorname{Im} E_{\sigma} = \frac{\sigma}{2}\gamma^{2}\Gamma_V+O(\gamma^3), \quad
	\operatorname{Im} k_{\sigma} = \frac{\gamma^{2}\Gamma_V}{2 v_B}+O(\gamma^3).
	\label{eq:Im_PT}
\end{equation}
For $\Gamma_V>0$, both poles ascend into the upper half of the complex $k$ plane [Fig.~\ref{fig:pole_trajectory}].
The $\sigma = +$ pole enters the first quadrant and becomes a TGBS whose intensity grows as $e^{\gamma^{2}\Gamma_V t}$, while the $\sigma = -$ pole enters the second quadrant and becomes a time-decaying bound state.
Since the TGBS emerges for arbitrarily small $\gamma$, the system is dynamically unstable with a strictly zero threshold.
Equations~\eqref{eq:Im_PT} and~\eqref{eq:Gamma} constitute the central result of this work.

This thresholdless instability rests on two independent ingredients, and the transverse $\mathcal{PT}$-symmetric configuration supplies both simultaneously.
First, the unperturbed pole must sit exactly on the real-$k$ axis.
A generic resonance carries a finite radiative width and its pole lies in the lower half-plane; lifting it across the real axis requires sufficient gain to compensate the radiative loss, setting a finite threshold.
A BIC carries exactly zero radiative width, so its pole starts on the real axis with no leakage to overcome.
Second, the perturbation must displace the pole upward rather than downward.
A Hermitian perturbation couples the BIC to the continuum and pushes the pole into the lower half-plane, as in Eq.~\eqref{eq:Im_Hermitian}.
The anti-Hermitian character of the gain-loss potential reverses the sign of the second-order shift through $\eta^{2}=-\gamma^{2}$, converting radiative decay into exponential growth.
The combination of a pole on the real axis and a strictly upward displacement guarantees that an infinitesimal $\gamma$ suffices to trigger instability.

The analysis above assumes $\Gamma_V>0$, that is, a nonvanishing on-shell projection of $V|B\rangle$ onto the propagating channels at energy $E_B$ [Eq.~\eqref{eq:Gamma}].
Since both $V$ and $|B\rangle$ are parity-odd, the state $V|B\rangle$ is parity-even and lies in the same symmetry sector as the scattering states, so the matrix elements $\langle\psi_{\pm k_B}|V|B\rangle$ are symmetry-allowed and generically nonzero.
The most direct route to restore a finite threshold is to enforce $V|B\rangle = 0$, which eliminates all corrections in Eq.~\eqref{eq:exact} and leaves the BIC intact under the perturbation.
This can be achieved when the gain-loss potential has zero spatial overlap with the BIC wave function, as demonstrated in Sec.~\ref{sec:examples}.

%%%%%%%%%%%%%%%%%%%%%%%%%%%%%%%%%%%%%%%%%%%%%%%%%%%%%%%%%%%%%%%%%%%%%%%%%%%%%%%%%%%%%%%%%%%%%%%%%%%%
\section{Illustrative examples}
\label{sec:examples}
%%%%%%%%%%%%%%%%%%%%%%%%%%%%%%%%%%%%%%%%%%%%%%%%%%%%%%%%%%%%%%%%%%%%%%%%%%%%%%%%%%%%%%%%%%%%%%%%%%%%

We now illustrate the general formalism of Sec.~\ref{sec:formalism} with two concrete models.
The three-site side-coupled model in Sec.~\ref{subsec:three_site} exhibits thresholdless instability despite possessing a unitary $S$ matrix at all gain-loss strengths.
The four-site rhombic model in Sec.~\ref{subsec:four_site} demonstrates that the spatial placement of the gain-loss potential relative to the BIC wave function controls whether the instability threshold vanishes or remains finite.

%%%%%%%%%%%%%%%%%%%%%%%%%%%%%%%%%%%%%%%%%%%%%%%%%%%%%%%%%%%%%%%%%%%%%%%%%%%%%%%%%%%%%%%%%%%%%%%%%%%%
\subsection{Three-site side-coupled model}
\label{subsec:three_site}
%%%%%%%%%%%%%%%%%%%%%%%%%%%%%%%%%%%%%%%%%%%%%%%%%%%%%%%%%%%%%%%%%%%%%%%%%%%%%%%%%%%%%%%%%%%%%%%%%%%%

\begin{figure}
\centering
\includegraphics{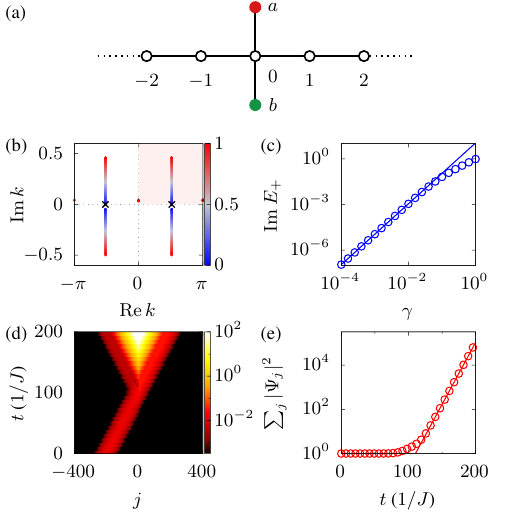}
\caption{
Thresholdless instability in the three-site side-coupled model.
(a) Schematic of the model with gain $+i\gamma$ on site $a$ and loss $-i\gamma$ on site $b$.
(b) Trajectories of the $S$-matrix poles in the complex $k$ plane as a function of $\gamma$, with the color bar showing $\gamma$ in units of $J$.
Black crosses mark the BIC at $k=\pm k_B$.
The pole originating from $+k_B$ enters the first quadrant (shaded) for any nonzero $\gamma$, signaling the emergence of a TGBS.
(c) Imaginary part of the TGBS energy as a function of $\gamma$ on a double-logarithmic scale.
Circles are exact pole solutions and the solid line is the perturbative prediction of Eq.~\eqref{eq:ImE3}.
(d) Time evolution of the intensity $|\Psi_j(t)|^{2}$ of the Gaussian wave packet with $k_0=\pi/2$, $\sigma=40$, and $j_0=-200$ on a lattice of $L=803$ sites for $\gamma=0.1J$.
(e) Total intensity $\sum_{j}|\Psi_j(t)|^{2}$ for the evolution in (d) on a semilogarithmic scale.
The solid line is a fit to $Ce^{2\alpha t}$ with $\alpha=(0.0646\pm0.0001)J$, in agreement with $\operatorname{Im}E_{+}$ of the exact pole.
Parameters: $g=J$, $\kappa=0.3J$, and $\epsilon_0=0.1J$.
}
\label{fig:three_site_model}
\end{figure}

We consider a scattering center composed of a chain site $0$ and two side-coupled sites $a$ and $b$ [Fig.~\ref{fig:three_site_model}(a)].
The left and right leads occupy sites $j\le-1$ and $j\ge1$, respectively.
This geometry with balanced gain and loss has been studied extensively as a $\mathcal{PT}$-symmetric Fano scatterer whose $S$ matrix remains unitary despite the non-Hermiticity of the scattering center~\cite{miroshnichenko2011a,jin2012,zhu2015,zhang2019c,ganguly2022}.
Its dynamical stability, however, has not been examined.

The Hamiltonian is $H=H_0+i\gamma V_3$, corresponding to Eq.~\eqref{eq:H_perturbed} with $\eta=i\gamma$.
The unperturbed part $H_0$ takes the form of Eq.~\eqref{eq:H0} with
\begin{equation}
H_{\mathrm{c}}=\epsilon_0\bigl(|a\rangle\langle a|+|b\rangle\langle b|\bigr)
-\kappa\bigl(|a\rangle\langle 0|+|b\rangle\langle 0|+\mathrm{H.c.}\bigr),
\label{eq:Hc3}
\end{equation}
\begin{equation}
H_{\mathrm{int}}=-g\bigl(|{-}1\rangle\langle 0|+|1\rangle\langle 0|+\mathrm{H.c.}\bigr),
\label{eq:Hint3}
\end{equation}
and the gain-loss potential is
\begin{equation}
V_3=|a\rangle\langle a|-|b\rangle\langle b|.
\label{eq:V3}
\end{equation}
The mirror operation $\mathcal{P}$ exchanges $a\leftrightarrow b$ and leaves all chain sites invariant, so that $[\mathcal{P},H_0]=0$ and $\mathcal{P}V_3\mathcal{P}^{-1}=-V_3$.

We first recall the stationary scattering properties.
The scattering state $|\psi_k\rangle$ satisfies $H|\psi_k\rangle=E|\psi_k\rangle$ with $E=-2J\cos k$.
Expanding $|\psi_k\rangle=\sum_j\psi(j)|j\rangle+\psi(a)|a\rangle+\psi(b)|b\rangle$ for a wave incident from the left [$k\in(0,\pi)$], the lead amplitudes take the form
\begin{equation}
\psi(j)=
\begin{cases}
e^{ikj}+re^{-ikj}, & j\le -1,\\
te^{ikj}, & j\ge 1,
\end{cases}
\label{eq:scatt_form}
\end{equation}
where $r$ and $t$ are the reflection and transmission amplitudes.
The eigenvalue equations at sites $a$ and $b$ give $\psi(a)=-\kappa\psi(0)/(E-\epsilon_0-i\gamma)$ and $\psi(b)=-\kappa\psi(0)/(E-\epsilon_0+i\gamma)$.
Substituting these into the eigenvalue equation at site $0$ yields
\begin{equation}
-g\bigl[\psi(-1)+\psi(1)\bigr] + \epsilon_{\mathrm{eff}}\,\psi(0) = E\,\psi(0),
\label{eq:eff_site0}
\end{equation}
with the effective onsite potential
\begin{equation}
\epsilon_{\mathrm{eff}} = \frac{2\kappa^2 (E-\epsilon_0)}{(E-\epsilon_0)^2+\gamma^2}.
\label{eq:eps_eff}
\end{equation}
Crucially, although the scattering center is non-Hermitian, $\epsilon_{\mathrm{eff}}$ is strictly real for all real $E$.
Evaluating the eigenvalue equation at $j=\pm1$ with the dispersion relation~\eqref{eq:dispersion} gives $g\psi(0)=J(1+r)$ and $g\psi(0)=Jt$, hence $1+r=t$.
Inserting $\psi(\pm1)$ from Eq.~\eqref{eq:scatt_form} and $\psi(0)=(J/g)t$ into Eq.~\eqref{eq:eff_site0}, we obtain
\begin{equation}
t = \frac{i\Lambda}{\delta + i\Lambda},
\quad
r = \frac{-\delta}{\delta + i\Lambda},
\label{eq:tr3}
\end{equation}
with
\begin{equation}
\delta = E\Bigl(1-\frac{g^2}{J^2}\Bigr) - \epsilon_{\mathrm{eff}},
\quad
\Lambda = \frac{2g^2}{J}\sin k.
\label{eq:delta_Lambda}
\end{equation}
Since $\delta$ and $\Lambda$ are both real, $|t|^{2}+|r|^{2}=1$ holds for all $\gamma$, confirming the unitarity of the $S$ matrix~\cite{jin2012,zhu2015}.
At $E = \epsilon_0$ with $\gamma = 0$, the potential $\epsilon_{\mathrm{eff}}$ diverges, forcing $\psi(0) = 0$ and yielding perfect reflection ($t = 0$, $r = -1$)~\cite{miroshnichenko2005a,miroshnichenko2009a}.
Any nonzero $\gamma$ regularizes $\epsilon_{\mathrm{eff}}$ to zero at this energy, giving perfect transmission ($t=1$, $r=0$) for uniform coupling $g=J$~\cite{miroshnichenko2011a,zhu2015}.
As we now show, this unitary stationary behavior masks a thresholdless dynamical instability triggered by the BIC.

At $\gamma=0$, the scattering center supports the parity-odd eigenstate 
\begin{equation}
|B_3\rangle=\frac{1}{\sqrt{2}}\bigl(|a\rangle-|b\rangle\bigr)
\label{eq:B3}
\end{equation}
at energy $E_B=\epsilon_0$.
For $|\epsilon_0|<2J$, this energy lies inside the continuum at the momenta $\pm k_B$ defined by $\epsilon_0=-2J\cos k_B$.
By the parity-mismatch mechanism of Sec.~\ref{sec:formalism}, $|B_3\rangle$ is decoupled from all scattering states and constitutes a symmetry-protected BIC.

Turning on $\gamma$ breaks the mirror symmetry that protects the BIC.
Since $V_3$ and $|B_3\rangle$ are both parity-odd, the state $V_3|B_3\rangle=(|a\rangle+|b\rangle)/\sqrt{2}$ is parity-even and thus couples to the scattering continuum.
Evaluating the Fermi-golden-rule coupling $\Gamma_V$ in Eq.~\eqref{eq:Gamma} requires the scattering states of $H_0$ at $E=E_B$.
Because $H_0$ exhibits perfect reflection at this energy, the left-incident state takes the form
\begin{equation}
\psi_{k_B}(j)=
\begin{cases}
2i\sin(k_B j), & j\le -1,\\[2pt]
0, & j\ge 0,
\end{cases}
\label{eq:psi_kB3}
\end{equation}
with side-site amplitudes $\psi_{k_B}(a)=\psi_{k_B}(b)=i(g/\kappa)\sin k_B$.
The right-incident state $\psi_{-k_B}$ follows from the left-right symmetry of $H_0$.
Each matrix element evaluates to $|\langle\psi_{\pm k_B}|V_3|B_3\rangle|=\sqrt{2}\,(g/\kappa)\sin k_B$, and with $v_B=2J\sin k_B$, Eq.~\eqref{eq:Gamma} yields
\begin{equation}
\Gamma_V=\frac{2g^{2}\sin k_B}{J\kappa^{2}}.
\label{eq:Gamma3}
\end{equation}
According to Eq.~\eqref{eq:Im_PT}, the pole emerging from $+k_B$ becomes a TGBS with growth rate
\begin{equation}
\operatorname{Im}E_{+} = \frac{g^{2}\sin k_B}{J\kappa^{2}}\,\gamma^{2} + O(\gamma^{3}),
\label{eq:ImE3}
\end{equation}
so the system is dynamically unstable for arbitrarily small $\gamma$.

To verify this prediction, we obtain the exact $S$-matrix poles by solving $\delta+i\Lambda=0$ numerically in the complex $k$ plane.
Figure~\ref{fig:three_site_model}(b) shows the pole trajectories as functions of $\gamma$ for $g=J$, $\kappa=0.3J$, and $\epsilon_0=0.1J$.
The pole pair originating from the BIC at $\pm k_B$ moves into the upper half-plane as soon as $\gamma$ is turned on, producing a TGBS in the first quadrant and a time-decaying bound state in the second, in accordance with the general scenario of Fig.~\ref{fig:pole_trajectory}.
Figure~\ref{fig:three_site_model}(c) compares the exact $\operatorname{Im}E_{+}$ with Eq.~\eqref{eq:ImE3}.
The quadratic law is followed closely at small $\gamma$, with higher-order corrections becoming visible only at large $\gamma$.

Finally, we demonstrate the instability in real time by simulating the evolution of a Gaussian wave packet.
The initial state is
\begin{equation}
|\Psi(0)\rangle=\mathcal{N}\sum_j e^{-(j-j_0)^{2}/2\sigma^{2}}e^{ik_0 j}|j\rangle,
\label{eq:wavepacket3}
\end{equation}
where $\mathcal{N}$ is a normalization constant, $j_0=-200$ is the initial position in the left lead, $\sigma=40$ controls the width, and $k_0=\pi/2$ is the central wave number.
The evolution $|\Psi(t)\rangle=e^{-iHt}|\Psi(0)\rangle$ is computed by exact diagonalization on a lattice of $L=803$ sites at $\gamma=0.1J$.
As shown in Fig.~\ref{fig:three_site_model}(d), the wave packet is partially transmitted and reflected, while a localized component at the scattering center grows exponentially and eventually dominates.
The total intensity in Fig.~\ref{fig:three_site_model}(d) follows $Ce^{2\alpha t}$ with fitted $\alpha=(0.0646\pm0.0001)J$, in excellent agreement with the exact pole value $\operatorname{Im}E_{+}\approx0.0646J$ at $\gamma=0.1J$.
These results confirm the thresholdless instability predicted in Sec.~\ref{subsec:instability}.

%%%%%%%%%%%%%%%%%%%%%%%%%%%%%%%%%%%%%%%%%%%%%%%%%%%%%%%%%%%%%%%%%%%%%%%%%%%%%%%%
\subsection{Four-site rhombic model}
\label{subsec:four_site}
%%%%%%%%%%%%%%%%%%%%%%%%%%%%%%%%%%%%%%%%%%%%%%%%%%%%%%%%%%%%%%%%%%%%%%%%%%%%%%%%

\begin{figure}
\centering
\includegraphics{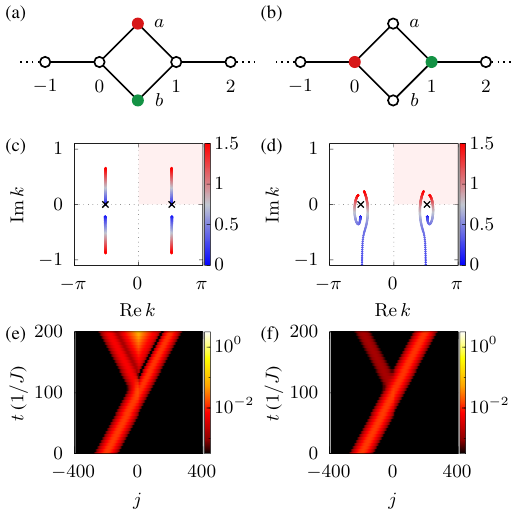}
\caption{
Placement-controlled stability in the four-site rhombic model.
(a),(b) Schematics of the transverse ($V_v$) and longitudinal ($V_h$) placements, respectively.
Gain and loss reside on the side sites $a,b$ in (a) and on the chain sites $0,1$ in (b).
(c),(d) Trajectories of the $S$-matrix poles in the complex $k$ plane as a function of $\gamma$ for the transverse and longitudinal placements, respectively; the color bar indicates $\gamma$ in units of $J$.
Black crosses mark the BIC at $\pm k_B$.
In (c), the pole pair originating from the BIC moves into the upper half-plane for arbitrarily small $\gamma$, signaling the emergence of a TGBS, whereas in (d) the BIC remains pinned on the real axis for all $\gamma$.
(e),(f) Time evolution of the wave-packet intensity $|\Psi_j(t)|^{2}$ at $\gamma=0.1J$ for the transverse and longitudinal placements, respectively.
The initial state is a Gaussian wave packet with $k_0=\pi/2$, $\sigma=40$, and $j_0=-200$ on a lattice of $L=802$ sites.
Exponential growth of the localized component appears in (e) but is absent in (f).
Parameters: $g=J$, $\kappa=0.3J$, and $\epsilon_0=0.1J$.
}
\label{fig:four_site_model}
\end{figure}

We next consider a rhombic scattering center composed of the chain sites $0$ and $1$ and two side sites $a$ and $b$ [Figs.~\ref{fig:four_site_model}(a) and \ref{fig:four_site_model}(b)]~\cite{jin2016a}.
The left and right leads occupy sites $j\le-1$ and $j\ge2$, respectively.
The unperturbed Hamiltonian $H_0$ takes the form of Eq.~\eqref{eq:H0} with
\begin{equation}
H_{\mathrm{c}}=\epsilon_0\bigl(|a\rangle\langle a|+|b\rangle\langle b|\bigr)
-\kappa\sum_{n=0,1}\bigl(|a\rangle\langle n|+|b\rangle\langle n|+\mathrm{H.c.}\bigr),
\label{eq:Hc4}
\end{equation}
\begin{equation}
H_{\mathrm{int}}=-g\bigl(|{-}1\rangle\langle 0|+|2\rangle\langle 1|+\mathrm{H.c.}\bigr).
\label{eq:Hint4}
\end{equation}
The mirror operation $\mathcal{P}$ exchanges $a\leftrightarrow b$ and leaves all chain sites invariant.
The parity-odd state
\begin{equation}
|B_4\rangle=\frac{1}{\sqrt{2}}\bigl(|a\rangle-|b\rangle\bigr)
\label{eq:B4}
\end{equation}
satisfies $H_0|B_4\rangle=\epsilon_0|B_4\rangle$ and constitutes a symmetry-protected BIC for $|\epsilon_0|<2J$.

The rhombic geometry admits two inequivalent placements of the balanced gain-loss potential,
\begin{equation}
V_v=|a\rangle\langle a|-|b\rangle\langle b|,\quad
V_h=|0\rangle\langle 0|-|1\rangle\langle 1|.
\label{eq:V4}
\end{equation}
The potential $i\gamma V_v$ places gain and loss on opposite sides of the transport axis, realizing a transverse $\mathcal{PT}$-symmetric configuration [Fig.~\ref{fig:four_site_model}(a)].
The potential $i\gamma V_h$ places them along the transport axis, realizing a longitudinal $\mathcal{PT}$-symmetric configuration [Fig.~\ref{fig:four_site_model}(b)].
We now show that these two placements lead to qualitatively different stability properties.

The transverse placement reproduces the scenario of the three-site side-coupled model.
The operator $V_v$ breaks the mirror symmetry and couples the BIC to the continuum through $V_v|B_4\rangle=(|a\rangle+|b\rangle)/\sqrt{2}$, which is parity-even.
A calculation analogous to that of Sec.~\ref{subsec:three_site} yields
\begin{equation}
\Gamma_{V}=\frac{g^{2}\sin k_B}{2J\kappa^{2}}.
\label{eq:Gamma4v}
\end{equation}
By Eq.~\eqref{eq:Im_PT}, a TGBS emerges with $\operatorname{Im}E_{+}=\gamma^{2}\Gamma_{V}/2+O(\gamma^{3})$ for arbitrarily small $\gamma$.
The pole trajectories in Fig.~\ref{fig:four_site_model}(c) confirm that the pole pair originating from the BIC enters the upper half of the complex $k$ plane as soon as $\gamma$ is turned on.
Accordingly, the wave-packet simulation at $\gamma=0.1J$ in Fig.~\ref{fig:four_site_model}(e) exhibits exponential growth localized at the scattering center.

The longitudinal placement behaves in the opposite way.
Since $V_h$ acts only on the chain sites, it commutes with $\mathcal{P}$, and the protecting mirror symmetry survives for all $\gamma$.
Moreover, $|B_4\rangle$ has zero amplitude on sites $0$ and $1$, so $V_h|B_4\rangle=0$ and every correction in Eq.~\eqref{eq:exact} vanishes identically.
The BIC therefore remains an exact eigenstate of the non-Hermitian Hamiltonian at the real energy $\epsilon_0$, and the corresponding poles stay pinned on the real axis [Fig.~\ref{fig:four_site_model}(d)].
An instability can arise only when some other pole crosses into the first quadrant, which requires a finite gain strength.
At $\gamma=0.1J$ all remaining poles lie in the lower half-plane, and the wave packet undergoes ordinary scattering without any growth [Fig.~\ref{fig:four_site_model}(f)].

The two placements thus produce opposite dynamical outcomes for the same scattering center at the same gain strength.
The transverse configuration is destabilized by an infinitesimal $\gamma$, whereas the longitudinal configuration preserves the BIC and retains a finite instability threshold.
This contrast confirms the zero-overlap criterion of Sec.~\ref{subsec:instability} and identifies the spatial placement of gain and loss as a practical design principle for stable $\mathcal{PT}$-symmetric scattering devices.

%%%%%%%%%%%%%%%%%%%%%%%%%%%%%%%%%%%%%%%%%%%%%%%%%%%%%%%%%%%%%%%%%%%%%%%%%%%%%%%%
\section{Summary and discussion}
\label{sec:conclusions}
%%%%%%%%%%%%%%%%%%%%%%%%%%%%%%%%%%%%%%%%%%%%%%%%%%%%%%%%%%%%%%%%%%%%%%%%%%%%%%%%

We have uncovered a thresholdless dynamical instability of transverse $\mathcal{PT}$-symmetric scattering systems.
In the absence of gain and loss, the up-down mirror symmetry of the transverse geometry generically protects a BIC, which appears as the exact coalescence of an $S$-matrix pole and a zero on the real axis of the complex $k$ plane.
The transverse gain-loss potential breaks this protecting symmetry, and its anti-Hermitian character reverses the sign of the second-order energy shift relative to a Hermitian perturbation.
Instead of acquiring a Fermi-golden-rule decay width, the BIC turns into a TGBS with $\operatorname{Im}E=\gamma^{2}\Gamma_{V}/2+O(\gamma^{3})$, where $\Gamma_{V}$ is the golden-rule coupling between the BIC and the continuum.
The instability therefore sets in at arbitrarily small $\gamma$ whenever $\Gamma_{V}>0$.
Exact pole trajectories and wave-packet simulations in two tight-binding models confirm this scenario quantitatively.

The three-site side-coupled model demonstrates that a unitary $S$ matrix does not guarantee dynamical stability.
Unitarity is a property of the stationary scattering states at real energies, whereas stability is governed by the discrete bound states at complex energies.
Once a TGBS emerges, it grows exponentially and eventually overwhelms the scattering response, rendering the stationary description unphysical.
The dynamical stability of a $\mathcal{PT}$-symmetric scattering system must therefore be verified independently of its stationary properties.
The criterion $\Gamma_{V}>0$ provides such a test and can be evaluated entirely within the unperturbed Hermitian system.

The four-site rhombic model turns this stability analysis into a design principle.
The condition $\Gamma_{V}=0$ alone does not ensure a finite threshold, since it removes only the second-order correction while higher-order terms can still destabilize the system.
The stronger condition $V|B\rangle=0$ eliminates all corrections to the pole energy and protects the BIC exactly.
The four-site rhombic model realizes this condition through the longitudinal placement of the gain-loss potential, which has zero spatial overlap with the BIC wave function and thereby retains a finite instability threshold.
The spatial overlap between the gain-loss potential and the BIC wave function thus constitutes a practical design principle for stable $\mathcal{PT}$-symmetric scattering devices.

Two remarks on the observability of the instability are in order.
First, since the growth rate is quadratic in $\gamma$, the instability develops slowly at weak gain-loss strength and may remain undetected within a finite observation window.
On sufficiently long time scales, however, the exponential growth inevitably dominates and spoils the designed scattering response.
Second, our analysis assumes a linear model in which the TGBS grows without bound.
In realistic settings, gain saturation and other nonlinear effects regularize this growth, but they do not remove the instability of the underlying linear dynamics.

Finally, the mechanism relies only on the mirror symmetry of the underlying Hermitian structure and the anti-Hermitian character of the gain-loss potential.
It therefore applies to a broad class of transverse $\mathcal{PT}$-symmetric systems, including side-coupled defects, parallel quantum dots, quantum rings, and ladder structures.
Coupled optical waveguide arrays, in which both $\mathcal{PT}$-symmetric potentials and BICs have been realized~\cite{plotnik2011,weimann2017}, provide a natural platform for testing our predictions.

%%%%%%%%%%%%%%%%%%%%%%%%%%%%%%%%%%%%%%%%%%%%%%%%%%%%%%%%%%%%%%%%%%%%%%%%%%%%%%%%%%%%%%%%%%%%%%%%%%%%
\begin{acknowledgments}
This work was supported by the Natural Science Foundation of the Jiangsu Higher Education Institutions of China, Grant No. 22KJB140009.
\end{acknowledgments}
%%%%%%%%%%%%%%%%%%%%%%%%%%%%%%%%%%%%%%%%%%%%%%%%%%%%%%%%%%%%%%%%%%%%%%%%%%%%%%%%%%%%%%%%%%%%%%%%%%%%

\bibliography{MyLibrary.bib}

\begin{thebibliography}{50}%
\makeatletter
\providecommand \@ifxundefined [1]{%
 \@ifx{#1\undefined}
}%
\providecommand \@ifnum [1]{%
 \ifnum #1\expandafter \@firstoftwo
 \else \expandafter \@secondoftwo
 \fi
}%
\providecommand \@ifx [1]{%
 \ifx #1\expandafter \@firstoftwo
 \else \expandafter \@secondoftwo
 \fi
}%
\providecommand \natexlab [1]{#1}%
\providecommand \enquote  [1]{``#1''}%
\providecommand \bibnamefont  [1]{#1}%
\providecommand \bibfnamefont [1]{#1}%
\providecommand \citenamefont [1]{#1}%
\providecommand \href@noop [0]{\@secondoftwo}%
\providecommand \href [0]{\begingroup \@sanitize@url \@href}%
\providecommand \@href[1]{\@@startlink{#1}\@@href}%
\providecommand \@@href[1]{\endgroup#1\@@endlink}%
\providecommand \@sanitize@url [0]{\catcode `\\12\catcode `\$12\catcode
  `\&12\catcode `\#12\catcode `\^12\catcode `\_12\catcode `\%12\relax}%
\providecommand \@@startlink[1]{}%
\providecommand \@@endlink[0]{}%
\providecommand \url  [0]{\begingroup\@sanitize@url \@url }%
\providecommand \@url [1]{\endgroup\@href {#1}{\urlprefix }}%
\providecommand \urlprefix  [0]{URL }%
\providecommand \Eprint [0]{\href }%
\providecommand \doibase [0]{https://doi.org/}%
\providecommand \selectlanguage [0]{\@gobble}%
\providecommand \bibinfo  [0]{\@secondoftwo}%
\providecommand \bibfield  [0]{\@secondoftwo}%
\providecommand \translation [1]{[#1]}%
\providecommand \BibitemOpen [0]{}%
\providecommand \bibitemStop [0]{}%
\providecommand \bibitemNoStop [0]{.\EOS\space}%
\providecommand \EOS [0]{\spacefactor3000\relax}%
\providecommand \BibitemShut  [1]{\csname bibitem#1\endcsname}%
\let\auto@bib@innerbib\@empty
%</preamble>
\bibitem [{\citenamefont {Bender}\ and\ \citenamefont
  {Boettcher}(1998)}]{bender1998a}%
  \BibitemOpen
  \bibfield  {author} {\bibinfo {author} {\bibfnamefont {C.~M.}\ \bibnamefont
  {Bender}}\ and\ \bibinfo {author} {\bibfnamefont {S.}~\bibnamefont
  {Boettcher}},\ }\bibfield  {title} {\bibinfo {title} {Real spectra in
  non-{{Hermitian Hamiltonians}} having {$\mathcal{PT}$} symmetry},\ }\href
  {https://doi.org/10.1103/PhysRevLett.80.5243} {\bibfield  {journal} {\bibinfo
   {journal} {Phys. Rev. Lett.}\ }\textbf {\bibinfo {volume} {80}},\ \bibinfo
  {pages} {5243} (\bibinfo {year} {1998})}\BibitemShut {NoStop}%
\bibitem [{\citenamefont {Bender}\ and\ \citenamefont
  {Hook}(2024)}]{bender2024}%
  \BibitemOpen
  \bibfield  {author} {\bibinfo {author} {\bibfnamefont {C.~M.}\ \bibnamefont
  {Bender}}\ and\ \bibinfo {author} {\bibfnamefont {D.~W.}\ \bibnamefont
  {Hook}},\ }\bibfield  {title} {\bibinfo {title} {{$\mathcal{PT}$}-symmetric
  quantum mechanics},\ }\href {https://doi.org/10.1103/RevModPhys.96.045002}
  {\bibfield  {journal} {\bibinfo  {journal} {Rev. Mod. Phys.}\ }\textbf
  {\bibinfo {volume} {96}},\ \bibinfo {pages} {045002} (\bibinfo {year}
  {2024})}\BibitemShut {NoStop}%
\bibitem [{\citenamefont {Konotop}\ \emph {et~al.}(2016)\citenamefont
  {Konotop}, \citenamefont {Yang},\ and\ \citenamefont
  {Zezyulin}}]{konotop2016}%
  \BibitemOpen
  \bibfield  {author} {\bibinfo {author} {\bibfnamefont {V.~V.}\ \bibnamefont
  {Konotop}}, \bibinfo {author} {\bibfnamefont {J.}~\bibnamefont {Yang}},\ and\
  \bibinfo {author} {\bibfnamefont {D.~A.}\ \bibnamefont {Zezyulin}},\
  }\bibfield  {title} {\bibinfo {title} {Nonlinear waves in
  {$\mathcal{PT}$}-symmetric systems},\ }\href
  {https://doi.org/10.1103/RevModPhys.88.035002} {\bibfield  {journal}
  {\bibinfo  {journal} {Rev. Mod. Phys.}\ }\textbf {\bibinfo {volume} {88}},\
  \bibinfo {pages} {035002} (\bibinfo {year} {2016})}\BibitemShut {NoStop}%
\bibitem [{\citenamefont {Feng}\ \emph {et~al.}(2017)\citenamefont {Feng},
  \citenamefont {{El-Ganainy}},\ and\ \citenamefont {Ge}}]{feng2017}%
  \BibitemOpen
  \bibfield  {author} {\bibinfo {author} {\bibfnamefont {L.}~\bibnamefont
  {Feng}}, \bibinfo {author} {\bibfnamefont {R.}~\bibnamefont {{El-Ganainy}}},\
  and\ \bibinfo {author} {\bibfnamefont {L.}~\bibnamefont {Ge}},\ }\bibfield
  {title} {\bibinfo {title} {Non-{{Hermitian}} photonics based on parity--time
  symmetry},\ }\href {https://doi.org/10.1038/s41566-017-0031-1} {\bibfield
  {journal} {\bibinfo  {journal} {Nat. Photonics}\ }\textbf {\bibinfo {volume}
  {11}},\ \bibinfo {pages} {752} (\bibinfo {year} {2017})}\BibitemShut
  {NoStop}%
\bibitem [{\citenamefont {{El-Ganainy}}\ \emph {et~al.}(2018)\citenamefont
  {{El-Ganainy}}, \citenamefont {Makris}, \citenamefont {Khajavikhan},
  \citenamefont {Musslimani}, \citenamefont {Rotter},\ and\ \citenamefont
  {Christodoulides}}]{el-ganainy2018}%
  \BibitemOpen
  \bibfield  {author} {\bibinfo {author} {\bibfnamefont {R.}~\bibnamefont
  {{El-Ganainy}}}, \bibinfo {author} {\bibfnamefont {K.~G.}\ \bibnamefont
  {Makris}}, \bibinfo {author} {\bibfnamefont {M.}~\bibnamefont {Khajavikhan}},
  \bibinfo {author} {\bibfnamefont {Z.~H.}\ \bibnamefont {Musslimani}},
  \bibinfo {author} {\bibfnamefont {S.}~\bibnamefont {Rotter}},\ and\ \bibinfo
  {author} {\bibfnamefont {D.~N.}\ \bibnamefont {Christodoulides}},\ }\bibfield
   {title} {\bibinfo {title} {Non-{{Hermitian}} physics and {{PT}} symmetry},\
  }\href {https://doi.org/10.1038/nphys4323} {\bibfield  {journal} {\bibinfo
  {journal} {Nat. Phys.}\ }\textbf {\bibinfo {volume} {14}},\ \bibinfo {pages}
  {11} (\bibinfo {year} {2018})}\BibitemShut {NoStop}%
\bibitem [{\citenamefont {{\"O}zdemir}\ \emph {et~al.}(2019)\citenamefont
  {{\"O}zdemir}, \citenamefont {Rotter}, \citenamefont {Nori},\ and\
  \citenamefont {Yang}}]{ozdemir2019}%
  \BibitemOpen
  \bibfield  {author} {\bibinfo {author} {\bibfnamefont {{\c S}.~K.}\
  \bibnamefont {{\"O}zdemir}}, \bibinfo {author} {\bibfnamefont
  {S.}~\bibnamefont {Rotter}}, \bibinfo {author} {\bibfnamefont
  {F.}~\bibnamefont {Nori}},\ and\ \bibinfo {author} {\bibfnamefont
  {L.}~\bibnamefont {Yang}},\ }\bibfield  {title} {\bibinfo {title}
  {Parity--time symmetry and exceptional points in photonics},\ }\href
  {https://doi.org/10.1038/s41563-019-0304-9} {\bibfield  {journal} {\bibinfo
  {journal} {Nat. Mater.}\ }\textbf {\bibinfo {volume} {18}},\ \bibinfo {pages}
  {783} (\bibinfo {year} {2019})}\BibitemShut {NoStop}%
\bibitem [{\citenamefont {Jin}\ \emph {et~al.}(2016)\citenamefont {Jin},
  \citenamefont {Zhang}, \citenamefont {Zhang},\ and\ \citenamefont
  {Song}}]{jin2016a}%
  \BibitemOpen
  \bibfield  {author} {\bibinfo {author} {\bibfnamefont {L.}~\bibnamefont
  {Jin}}, \bibinfo {author} {\bibfnamefont {X.~Z.}\ \bibnamefont {Zhang}},
  \bibinfo {author} {\bibfnamefont {G.}~\bibnamefont {Zhang}},\ and\ \bibinfo
  {author} {\bibfnamefont {Z.}~\bibnamefont {Song}},\ }\bibfield  {title}
  {\bibinfo {title} {Reciprocal and unidirectional scattering of parity-time
  symmetric structures},\ }\href {https://doi.org/10.1038/srep20976} {\bibfield
   {journal} {\bibinfo  {journal} {Sci. Rep.}\ }\textbf {\bibinfo {volume}
  {6}},\ \bibinfo {pages} {20976} (\bibinfo {year} {2016})}\BibitemShut
  {NoStop}%
\bibitem [{\citenamefont {Ramezani}\ \emph {et~al.}(2014)\citenamefont
  {Ramezani}, \citenamefont {Li}, \citenamefont {Wang},\ and\ \citenamefont
  {Zhang}}]{ramezani2014}%
  \BibitemOpen
  \bibfield  {author} {\bibinfo {author} {\bibfnamefont {H.}~\bibnamefont
  {Ramezani}}, \bibinfo {author} {\bibfnamefont {H.-K.}\ \bibnamefont {Li}},
  \bibinfo {author} {\bibfnamefont {Y.}~\bibnamefont {Wang}},\ and\ \bibinfo
  {author} {\bibfnamefont {X.}~\bibnamefont {Zhang}},\ }\bibfield  {title}
  {\bibinfo {title} {Unidirectional spectral singularities},\ }\href
  {https://doi.org/10.1103/PhysRevLett.113.263905} {\bibfield  {journal}
  {\bibinfo  {journal} {Phys. Rev. Lett.}\ }\textbf {\bibinfo {volume} {113}},\
  \bibinfo {pages} {263905} (\bibinfo {year} {2014})}\BibitemShut {NoStop}%
\bibitem [{\citenamefont {Garmon}\ \emph {et~al.}(2015)\citenamefont {Garmon},
  \citenamefont {Gianfreda},\ and\ \citenamefont {Hatano}}]{garmon2015}%
  \BibitemOpen
  \bibfield  {author} {\bibinfo {author} {\bibfnamefont {S.}~\bibnamefont
  {Garmon}}, \bibinfo {author} {\bibfnamefont {M.}~\bibnamefont {Gianfreda}},\
  and\ \bibinfo {author} {\bibfnamefont {N.}~\bibnamefont {Hatano}},\
  }\bibfield  {title} {\bibinfo {title} {Bound states, scattering states, and
  resonant states in {$\mathcal{PT}$}-symmetric open quantum systems},\ }\href
  {https://doi.org/10.1103/PhysRevA.92.022125} {\bibfield  {journal} {\bibinfo
  {journal} {Phys. Rev. A}\ }\textbf {\bibinfo {volume} {92}},\ \bibinfo
  {pages} {022125} (\bibinfo {year} {2015})}\BibitemShut {NoStop}%
\bibitem [{\citenamefont {Zhu}\ \emph {et~al.}(2016)\citenamefont {Zhu},
  \citenamefont {L{\"u}},\ and\ \citenamefont {Chen}}]{zhu2016}%
  \BibitemOpen
  \bibfield  {author} {\bibinfo {author} {\bibfnamefont {B.}~\bibnamefont
  {Zhu}}, \bibinfo {author} {\bibfnamefont {R.}~\bibnamefont {L{\"u}}},\ and\
  \bibinfo {author} {\bibfnamefont {S.}~\bibnamefont {Chen}},\ }\bibfield
  {title} {\bibinfo {title} {{$\mathcal{PT}$}-symmetry breaking for the
  scattering problem in a one-dimensional non-{{Hermitian}} lattice model},\
  }\href {https://doi.org/10.1103/PhysRevA.93.032129} {\bibfield  {journal}
  {\bibinfo  {journal} {Phys. Rev. A}\ }\textbf {\bibinfo {volume} {93}},\
  \bibinfo {pages} {032129} (\bibinfo {year} {2016})}\BibitemShut {NoStop}%
\bibitem [{\citenamefont {Achilleos}\ \emph {et~al.}(2017)\citenamefont
  {Achilleos}, \citenamefont {Aur{\'e}gan},\ and\ \citenamefont
  {Pagneux}}]{achilleos2017a}%
  \BibitemOpen
  \bibfield  {author} {\bibinfo {author} {\bibfnamefont {V.}~\bibnamefont
  {Achilleos}}, \bibinfo {author} {\bibfnamefont {Y.}~\bibnamefont
  {Aur{\'e}gan}},\ and\ \bibinfo {author} {\bibfnamefont {V.}~\bibnamefont
  {Pagneux}},\ }\bibfield  {title} {\bibinfo {title} {Scattering by finite
  periodic {$\mathcal{P}\mathcal{T}$}-symmetric structures},\ }\href
  {https://doi.org/10.1103/PhysRevLett.119.243904} {\bibfield  {journal}
  {\bibinfo  {journal} {Phys. Rev. Lett.}\ }\textbf {\bibinfo {volume} {119}},\
  \bibinfo {pages} {243904} (\bibinfo {year} {2017})}\BibitemShut {NoStop}%
\bibitem [{\citenamefont {Shobe}\ \emph {et~al.}(2021)\citenamefont {Shobe},
  \citenamefont {Kuramoto}, \citenamefont {Imura},\ and\ \citenamefont
  {Hatano}}]{shobe2021}%
  \BibitemOpen
  \bibfield  {author} {\bibinfo {author} {\bibfnamefont {K.}~\bibnamefont
  {Shobe}}, \bibinfo {author} {\bibfnamefont {K.}~\bibnamefont {Kuramoto}},
  \bibinfo {author} {\bibfnamefont {K.-I.}\ \bibnamefont {Imura}},\ and\
  \bibinfo {author} {\bibfnamefont {N.}~\bibnamefont {Hatano}},\ }\bibfield
  {title} {\bibinfo {title} {Non-{{Hermitian Fabry-P\'erot}} resonances in a
  {$PT$}-symmetric system},\ }\href
  {https://doi.org/10.1103/PhysRevResearch.3.013223} {\bibfield  {journal}
  {\bibinfo  {journal} {Phys. Rev. Res.}\ }\textbf {\bibinfo {volume} {3}},\
  \bibinfo {pages} {013223} (\bibinfo {year} {2021})}\BibitemShut {NoStop}%
\bibitem [{\citenamefont {Ge}\ \emph {et~al.}(2012)\citenamefont {Ge},
  \citenamefont {Chong},\ and\ \citenamefont {Stone}}]{ge2012}%
  \BibitemOpen
  \bibfield  {author} {\bibinfo {author} {\bibfnamefont {L.}~\bibnamefont
  {Ge}}, \bibinfo {author} {\bibfnamefont {Y.~D.}\ \bibnamefont {Chong}},\ and\
  \bibinfo {author} {\bibfnamefont {A.~D.}\ \bibnamefont {Stone}},\ }\bibfield
  {title} {\bibinfo {title} {Conservation relations and anisotropic
  transmission resonances in one-dimensional {$\mathcal{PT}$}-symmetric
  photonic heterostructures},\ }\href
  {https://doi.org/10.1103/PhysRevA.85.023802} {\bibfield  {journal} {\bibinfo
  {journal} {Phys. Rev. A}\ }\textbf {\bibinfo {volume} {85}},\ \bibinfo
  {pages} {023802} (\bibinfo {year} {2012})}\BibitemShut {NoStop}%
\bibitem [{\citenamefont {Lin}\ \emph {et~al.}(2011)\citenamefont {Lin},
  \citenamefont {Ramezani}, \citenamefont {Eichelkraut}, \citenamefont
  {Kottos}, \citenamefont {Cao},\ and\ \citenamefont
  {Christodoulides}}]{lin2011}%
  \BibitemOpen
  \bibfield  {author} {\bibinfo {author} {\bibfnamefont {Z.}~\bibnamefont
  {Lin}}, \bibinfo {author} {\bibfnamefont {H.}~\bibnamefont {Ramezani}},
  \bibinfo {author} {\bibfnamefont {T.}~\bibnamefont {Eichelkraut}}, \bibinfo
  {author} {\bibfnamefont {T.}~\bibnamefont {Kottos}}, \bibinfo {author}
  {\bibfnamefont {H.}~\bibnamefont {Cao}},\ and\ \bibinfo {author}
  {\bibfnamefont {D.~N.}\ \bibnamefont {Christodoulides}},\ }\bibfield  {title}
  {\bibinfo {title} {Unidirectional invisibility induced by
  {$\mathcal{P}\mathcal{T}$}-symmetric periodic structures},\ }\href
  {https://doi.org/10.1103/PhysRevLett.106.213901} {\bibfield  {journal}
  {\bibinfo  {journal} {Phys. Rev. Lett.}\ }\textbf {\bibinfo {volume} {106}},\
  \bibinfo {pages} {213901} (\bibinfo {year} {2011})}\BibitemShut {NoStop}%
\bibitem [{\citenamefont {Feng}\ \emph {et~al.}(2013)\citenamefont {Feng},
  \citenamefont {Xu}, \citenamefont {Fegadolli}, \citenamefont {Lu},
  \citenamefont {Oliveira}, \citenamefont {Almeida}, \citenamefont {Chen},\
  and\ \citenamefont {Scherer}}]{feng2013}%
  \BibitemOpen
  \bibfield  {author} {\bibinfo {author} {\bibfnamefont {L.}~\bibnamefont
  {Feng}}, \bibinfo {author} {\bibfnamefont {Y.-L.}\ \bibnamefont {Xu}},
  \bibinfo {author} {\bibfnamefont {W.~S.}\ \bibnamefont {Fegadolli}}, \bibinfo
  {author} {\bibfnamefont {M.-H.}\ \bibnamefont {Lu}}, \bibinfo {author}
  {\bibfnamefont {J.~E.~B.}\ \bibnamefont {Oliveira}}, \bibinfo {author}
  {\bibfnamefont {V.~R.}\ \bibnamefont {Almeida}}, \bibinfo {author}
  {\bibfnamefont {Y.-F.}\ \bibnamefont {Chen}},\ and\ \bibinfo {author}
  {\bibfnamefont {A.}~\bibnamefont {Scherer}},\ }\bibfield  {title} {\bibinfo
  {title} {Experimental demonstration of a unidirectional reflectionless
  parity-time metamaterial at optical frequencies},\ }\href
  {https://doi.org/10.1038/nmat3495} {\bibfield  {journal} {\bibinfo  {journal}
  {Nat. Mater.}\ }\textbf {\bibinfo {volume} {12}},\ \bibinfo {pages} {108}
  (\bibinfo {year} {2013})}\BibitemShut {NoStop}%
\bibitem [{\citenamefont {Longhi}(2010)}]{longhi2010}%
  \BibitemOpen
  \bibfield  {author} {\bibinfo {author} {\bibfnamefont {S.}~\bibnamefont
  {Longhi}},\ }\bibfield  {title} {\bibinfo {title} {{$\mathcal{PT}$}-symmetric
  laser absorber},\ }\href {https://doi.org/10.1103/PhysRevA.82.031801}
  {\bibfield  {journal} {\bibinfo  {journal} {Phys. Rev. A}\ }\textbf {\bibinfo
  {volume} {82}},\ \bibinfo {pages} {031801} (\bibinfo {year}
  {2010})}\BibitemShut {NoStop}%
\bibitem [{\citenamefont {Chong}\ \emph {et~al.}(2011)\citenamefont {Chong},
  \citenamefont {Ge},\ and\ \citenamefont {Stone}}]{chong2011}%
  \BibitemOpen
  \bibfield  {author} {\bibinfo {author} {\bibfnamefont {Y.~D.}\ \bibnamefont
  {Chong}}, \bibinfo {author} {\bibfnamefont {L.}~\bibnamefont {Ge}},\ and\
  \bibinfo {author} {\bibfnamefont {A.~D.}\ \bibnamefont {Stone}},\ }\bibfield
  {title} {\bibinfo {title} {{$\mathcal{P}\mathcal{T}$}-symmetry breaking and
  laser-absorber modes in optical scattering systems},\ }\href
  {https://doi.org/10.1103/PhysRevLett.106.093902} {\bibfield  {journal}
  {\bibinfo  {journal} {Phys. Rev. Lett.}\ }\textbf {\bibinfo {volume} {106}},\
  \bibinfo {pages} {093902} (\bibinfo {year} {2011})}\BibitemShut {NoStop}%
\bibitem [{\citenamefont {Jin}\ and\ \citenamefont {Song}(2012)}]{jin2012}%
  \BibitemOpen
  \bibfield  {author} {\bibinfo {author} {\bibfnamefont {L.}~\bibnamefont
  {Jin}}\ and\ \bibinfo {author} {\bibfnamefont {Z.}~\bibnamefont {Song}},\
  }\bibfield  {title} {\bibinfo {title} {Hermitian scattering behavior for a
  non-{{Hermitian}} scattering center},\ }\href
  {https://doi.org/10.1103/PhysRevA.85.012111} {\bibfield  {journal} {\bibinfo
  {journal} {Phys. Rev. A}\ }\textbf {\bibinfo {volume} {85}},\ \bibinfo
  {pages} {012111} (\bibinfo {year} {2012})}\BibitemShut {NoStop}%
\bibitem [{\citenamefont {Zhu}\ \emph {et~al.}(2015)\citenamefont {Zhu},
  \citenamefont {L{\"u}},\ and\ \citenamefont {Chen}}]{zhu2015}%
  \BibitemOpen
  \bibfield  {author} {\bibinfo {author} {\bibfnamefont {B.}~\bibnamefont
  {Zhu}}, \bibinfo {author} {\bibfnamefont {R.}~\bibnamefont {L{\"u}}},\ and\
  \bibinfo {author} {\bibfnamefont {S.}~\bibnamefont {Chen}},\ }\bibfield
  {title} {\bibinfo {title} {Interplay between {{Fano}} resonance and
  {$\mathcal{PT}$} symmetry in non-{{Hermitian}} discrete systems},\ }\href
  {https://doi.org/10.1103/PhysRevA.91.042131} {\bibfield  {journal} {\bibinfo
  {journal} {Phys. Rev. A}\ }\textbf {\bibinfo {volume} {91}},\ \bibinfo
  {pages} {042131} (\bibinfo {year} {2015})}\BibitemShut {NoStop}%
\bibitem [{\citenamefont {Jin}(2022)}]{jin2022}%
  \BibitemOpen
  \bibfield  {author} {\bibinfo {author} {\bibfnamefont {L.}~\bibnamefont
  {Jin}},\ }\bibfield  {title} {\bibinfo {title} {Unitary {{Scattering
  Protected}} by {{Pseudo-Hermiticity}}},\ }\href
  {https://doi.org/10.1088/0256-307X/39/3/037302} {\bibfield  {journal}
  {\bibinfo  {journal} {Chin. Phys. Lett.}\ }\textbf {\bibinfo {volume} {39}},\
  \bibinfo {pages} {037302} (\bibinfo {year} {2022})}\BibitemShut {NoStop}%
\bibitem [{\citenamefont {Ganguly}\ \emph {et~al.}(2022)\citenamefont
  {Ganguly}, \citenamefont {Roy},\ and\ \citenamefont {Maiti}}]{ganguly2022}%
  \BibitemOpen
  \bibfield  {author} {\bibinfo {author} {\bibfnamefont {S.}~\bibnamefont
  {Ganguly}}, \bibinfo {author} {\bibfnamefont {S.}~\bibnamefont {Roy}},\ and\
  \bibinfo {author} {\bibfnamefont {S.~K.}\ \bibnamefont {Maiti}},\ }\bibfield
  {title} {\bibinfo {title} {Transport characteristics of a {${\mathcal
  {PT}}$}-symmetric non-{{Hermitian}} system: effect of environmental
  interaction},\ }\href {https://doi.org/10.1140/epjp/s13360-022-03016-8}
  {\bibfield  {journal} {\bibinfo  {journal} {Eur. Phys. J. Plus}\ }\textbf
  {\bibinfo {volume} {137}},\ \bibinfo {pages} {780} (\bibinfo {year}
  {2022})}\BibitemShut {NoStop}%
\bibitem [{\citenamefont {Soori}\ \emph {et~al.}(2023)\citenamefont {Soori},
  \citenamefont {Sivakumar},\ and\ \citenamefont {Subrahmanyam}}]{soori2023}%
  \BibitemOpen
  \bibfield  {author} {\bibinfo {author} {\bibfnamefont {A.}~\bibnamefont
  {Soori}}, \bibinfo {author} {\bibfnamefont {M.}~\bibnamefont {Sivakumar}},\
  and\ \bibinfo {author} {\bibfnamefont {V.}~\bibnamefont {Subrahmanyam}},\
  }\bibfield  {title} {\bibinfo {title} {Transmission across non-{{Hermitian}}
  {$\mathcal{PT}$}-symmetric quantum dots and ladders},\ }\href
  {https://doi.org/10.1088/1361-648X/aca3ec} {\bibfield  {journal} {\bibinfo
  {journal} {J. Phys.: Condens. Matter}\ }\textbf {\bibinfo {volume} {35}},\
  \bibinfo {pages} {055301} (\bibinfo {year} {2023})}\BibitemShut {NoStop}%
\bibitem [{\citenamefont {Miroshnichenko}\ \emph {et~al.}(2011)\citenamefont
  {Miroshnichenko}, \citenamefont {Malomed},\ and\ \citenamefont
  {Kivshar}}]{miroshnichenko2011a}%
  \BibitemOpen
  \bibfield  {author} {\bibinfo {author} {\bibfnamefont {A.~E.}\ \bibnamefont
  {Miroshnichenko}}, \bibinfo {author} {\bibfnamefont {B.~A.}\ \bibnamefont
  {Malomed}},\ and\ \bibinfo {author} {\bibfnamefont {Y.~S.}\ \bibnamefont
  {Kivshar}},\ }\bibfield  {title} {\bibinfo {title} {Nonlinearly
  {$\mathcal{PT}$}-symmetric systems: {{Spontaneous}} symmetry breaking and
  transmission resonances},\ }\href
  {https://doi.org/10.1103/PhysRevA.84.012123} {\bibfield  {journal} {\bibinfo
  {journal} {Phys. Rev. A}\ }\textbf {\bibinfo {volume} {84}},\ \bibinfo
  {pages} {012123} (\bibinfo {year} {2011})}\BibitemShut {NoStop}%
\bibitem [{\citenamefont {Zhang}\ \emph
  {et~al.}(2019{\natexlab{a}})\citenamefont {Zhang}, \citenamefont {Li},
  \citenamefont {Zhan}, \citenamefont {Yi},\ and\ \citenamefont
  {Gong}}]{zhang2019c}%
  \BibitemOpen
  \bibfield  {author} {\bibinfo {author} {\bibfnamefont {L.-L.}\ \bibnamefont
  {Zhang}}, \bibinfo {author} {\bibfnamefont {Z.-Z.}\ \bibnamefont {Li}},
  \bibinfo {author} {\bibfnamefont {G.-H.}\ \bibnamefont {Zhan}}, \bibinfo
  {author} {\bibfnamefont {G.-Y.}\ \bibnamefont {Yi}},\ and\ \bibinfo {author}
  {\bibfnamefont {W.-J.}\ \bibnamefont {Gong}},\ }\bibfield  {title} {\bibinfo
  {title} {Eigenenergies and quantum transport properties in a
  non-{{Hermitian}} quantum-dot chain with side-coupled dots},\ }\href
  {https://doi.org/10.1103/PhysRevA.99.032119} {\bibfield  {journal} {\bibinfo
  {journal} {Phys. Rev. A}\ }\textbf {\bibinfo {volume} {99}},\ \bibinfo
  {pages} {032119} (\bibinfo {year} {2019}{\natexlab{a}})}\BibitemShut
  {NoStop}%
\bibitem [{\citenamefont {Zhang}\ \emph
  {et~al.}(2019{\natexlab{b}})\citenamefont {Zhang}, \citenamefont {Zhan},
  \citenamefont {He}, \citenamefont {Zhang},\ and\ \citenamefont
  {Gong}}]{zhang2019d}%
  \BibitemOpen
  \bibfield  {author} {\bibinfo {author} {\bibfnamefont {L.-L.}\ \bibnamefont
  {Zhang}}, \bibinfo {author} {\bibfnamefont {G.-H.}\ \bibnamefont {Zhan}},
  \bibinfo {author} {\bibfnamefont {J.}~\bibnamefont {He}}, \bibinfo {author}
  {\bibfnamefont {Y.}~\bibnamefont {Zhang}},\ and\ \bibinfo {author}
  {\bibfnamefont {W.-J.}\ \bibnamefont {Gong}},\ }\bibfield  {title} {\bibinfo
  {title} {Transmission in a dimerized chain influenced by {${\mathcal
  {PT}}$}-symmetric potentials},\ }\href
  {https://doi.org/10.1088/1402-4896/ab1308} {\bibfield  {journal} {\bibinfo
  {journal} {Phys. Scr.}\ }\textbf {\bibinfo {volume} {94}},\ \bibinfo {pages}
  {085216} (\bibinfo {year} {2019}{\natexlab{b}})}\BibitemShut {NoStop}%
\bibitem [{\citenamefont {Li}\ \emph {et~al.}(2020)\citenamefont {Li},
  \citenamefont {Huang}, \citenamefont {He}, \citenamefont {Zhang},\ and\
  \citenamefont {Gong}}]{li2020a}%
  \BibitemOpen
  \bibfield  {author} {\bibinfo {author} {\bibfnamefont {X.-S.}\ \bibnamefont
  {Li}}, \bibinfo {author} {\bibfnamefont {P.-P.}\ \bibnamefont {Huang}},
  \bibinfo {author} {\bibfnamefont {J.}~\bibnamefont {He}}, \bibinfo {author}
  {\bibfnamefont {L.-L.}\ \bibnamefont {Zhang}},\ and\ \bibinfo {author}
  {\bibfnamefont {W.-J.}\ \bibnamefont {Gong}},\ }\bibfield  {title} {\bibinfo
  {title} {Fano effect in a one-dimensional photonic lattice with side-coupled
  {$\mathcal{PT}$}-symmetric non-{{Hermitian}} defects},\ }\href
  {https://doi.org/10.1364/OE.383301} {\bibfield  {journal} {\bibinfo
  {journal} {Opt. Express}\ }\textbf {\bibinfo {volume} {28}},\ \bibinfo
  {pages} {8560} (\bibinfo {year} {2020})}\BibitemShut {NoStop}%
\bibitem [{\citenamefont {Niu}\ \emph {et~al.}(2026)\citenamefont {Niu},
  \citenamefont {Xu}, \citenamefont {Liu}, \citenamefont {Yao},\ and\
  \citenamefont {Luo}}]{niu2026}%
  \BibitemOpen
  \bibfield  {author} {\bibinfo {author} {\bibfnamefont {P.}~\bibnamefont
  {Niu}}, \bibinfo {author} {\bibfnamefont {L.}~\bibnamefont {Xu}}, \bibinfo
  {author} {\bibfnamefont {H.}~\bibnamefont {Liu}}, \bibinfo {author}
  {\bibfnamefont {H.}~\bibnamefont {Yao}},\ and\ \bibinfo {author}
  {\bibfnamefont {H.-G.}\ \bibnamefont {Luo}},\ }\bibfield  {title} {\bibinfo
  {title} {Anomalous {{Fano}} effect in non-{{Hermitian}} double quantum
  dots},\ }\href {https://doi.org/10.1016/j.physb.2025.418139} {\bibfield
  {journal} {\bibinfo  {journal} {Phys. B (Amsterdam, Neth.)}\ }\textbf
  {\bibinfo {volume} {723}},\ \bibinfo {pages} {418139} (\bibinfo {year}
  {2026})}\BibitemShut {NoStop}%
\bibitem [{\citenamefont {Li}\ \emph {et~al.}(2017)\citenamefont {Li},
  \citenamefont {Jin},\ and\ \citenamefont {Song}}]{li2017}%
  \BibitemOpen
  \bibfield  {author} {\bibinfo {author} {\bibfnamefont {C.}~\bibnamefont
  {Li}}, \bibinfo {author} {\bibfnamefont {L.}~\bibnamefont {Jin}},\ and\
  \bibinfo {author} {\bibfnamefont {Z.}~\bibnamefont {Song}},\ }\bibfield
  {title} {\bibinfo {title} {Non-{{Hermitian}} interferometer:
  {{Unidirectional}} amplification without distortion},\ }\href
  {https://doi.org/10.1103/PhysRevA.95.022125} {\bibfield  {journal} {\bibinfo
  {journal} {Phys. Rev. A}\ }\textbf {\bibinfo {volume} {95}},\ \bibinfo
  {pages} {022125} (\bibinfo {year} {2017})}\BibitemShut {NoStop}%
\bibitem [{\citenamefont {Zhang}\ \emph {et~al.}(2017)\citenamefont {Zhang},
  \citenamefont {Zhan}, \citenamefont {Li},\ and\ \citenamefont
  {Gong}}]{zhang2017a}%
  \BibitemOpen
  \bibfield  {author} {\bibinfo {author} {\bibfnamefont {L.-L.}\ \bibnamefont
  {Zhang}}, \bibinfo {author} {\bibfnamefont {G.-H.}\ \bibnamefont {Zhan}},
  \bibinfo {author} {\bibfnamefont {Z.-Z.}\ \bibnamefont {Li}},\ and\ \bibinfo
  {author} {\bibfnamefont {W.-J.}\ \bibnamefont {Gong}},\ }\bibfield  {title}
  {\bibinfo {title} {Effect of {$\mathcal{PT}$} symmetry in a parallel
  double-quantum-dot structure},\ }\href
  {https://doi.org/10.1103/PhysRevA.96.062133} {\bibfield  {journal} {\bibinfo
  {journal} {Phys. Rev. A}\ }\textbf {\bibinfo {volume} {96}},\ \bibinfo
  {pages} {062133} (\bibinfo {year} {2017})}\BibitemShut {NoStop}%
\bibitem [{\citenamefont {Zhang}\ and\ \citenamefont
  {Gong}(2017)}]{zhang2017b}%
  \BibitemOpen
  \bibfield  {author} {\bibinfo {author} {\bibfnamefont {L.-L.}\ \bibnamefont
  {Zhang}}\ and\ \bibinfo {author} {\bibfnamefont {W.-J.}\ \bibnamefont
  {Gong}},\ }\bibfield  {title} {\bibinfo {title} {Transport properties in a
  non-{{Hermitian}} triple-quantum-dot structure},\ }\href
  {https://doi.org/10.1103/PhysRevA.95.062123} {\bibfield  {journal} {\bibinfo
  {journal} {Phys. Rev. A}\ }\textbf {\bibinfo {volume} {95}},\ \bibinfo
  {pages} {062123} (\bibinfo {year} {2017})}\BibitemShut {NoStop}%
\bibitem [{\citenamefont {Zhang}\ \emph {et~al.}(2020)\citenamefont {Zhang},
  \citenamefont {Gong}, \citenamefont {Yi},\ and\ \citenamefont
  {Du}}]{zhang2020e}%
  \BibitemOpen
  \bibfield  {author} {\bibinfo {author} {\bibfnamefont {L.-L.}\ \bibnamefont
  {Zhang}}, \bibinfo {author} {\bibfnamefont {W.-J.}\ \bibnamefont {Gong}},
  \bibinfo {author} {\bibfnamefont {G.-Y.}\ \bibnamefont {Yi}},\ and\ \bibinfo
  {author} {\bibfnamefont {A.}~\bibnamefont {Du}},\ }\bibfield  {title}
  {\bibinfo {title} {Influence of {$\mathcal{PT}$}-symmetric complex potentials
  on the decoupling mechanism in quantum transport processes},\ }\href
  {https://doi.org/10.1016/j.aop.2020.168162} {\bibfield  {journal} {\bibinfo
  {journal} {Ann. Phys.}\ }\textbf {\bibinfo {volume} {416}},\ \bibinfo {pages}
  {168162} (\bibinfo {year} {2020})}\BibitemShut {NoStop}%
\bibitem [{\citenamefont {Zheng}(2025)}]{zheng2025a}%
  \BibitemOpen
  \bibfield  {author} {\bibinfo {author} {\bibfnamefont {C.}~\bibnamefont
  {Zheng}},\ }\bibfield  {title} {\bibinfo {title} {Physical relevance of
  time-independent scattering calculations in non-{{Hermitian}} systems:
  {{The}} role of time-growing bound states},\ }\href
  {https://doi.org/10.1103/gzyf-77hr} {\bibfield  {journal} {\bibinfo
  {journal} {Phys. Rev. A}\ }\textbf {\bibinfo {volume} {112}},\ \bibinfo
  {pages} {042228} (\bibinfo {year} {2025})}\BibitemShut {NoStop}%
\bibitem [{\citenamefont {Bendix}\ \emph {et~al.}(2009)\citenamefont {Bendix},
  \citenamefont {Fleischmann}, \citenamefont {Kottos},\ and\ \citenamefont
  {Shapiro}}]{bendix2009}%
  \BibitemOpen
  \bibfield  {author} {\bibinfo {author} {\bibfnamefont {O.}~\bibnamefont
  {Bendix}}, \bibinfo {author} {\bibfnamefont {R.}~\bibnamefont {Fleischmann}},
  \bibinfo {author} {\bibfnamefont {T.}~\bibnamefont {Kottos}},\ and\ \bibinfo
  {author} {\bibfnamefont {B.}~\bibnamefont {Shapiro}},\ }\bibfield  {title}
  {\bibinfo {title} {Exponentially fragile {$\mathcal{PT}$} symmetry in
  lattices with localized eigenmodes},\ }\href
  {https://doi.org/10.1103/PhysRevLett.103.030402} {\bibfield  {journal}
  {\bibinfo  {journal} {Phys. Rev. Lett.}\ }\textbf {\bibinfo {volume} {103}},\
  \bibinfo {pages} {030402} (\bibinfo {year} {2009})}\BibitemShut {NoStop}%
\bibitem [{\citenamefont {Dmitriev}\ \emph {et~al.}(2011)\citenamefont
  {Dmitriev}, \citenamefont {Suchkov}, \citenamefont {Sukhorukov},\ and\
  \citenamefont {Kivshar}}]{dmitriev2011}%
  \BibitemOpen
  \bibfield  {author} {\bibinfo {author} {\bibfnamefont {S.~V.}\ \bibnamefont
  {Dmitriev}}, \bibinfo {author} {\bibfnamefont {S.~V.}\ \bibnamefont
  {Suchkov}}, \bibinfo {author} {\bibfnamefont {A.~A.}\ \bibnamefont
  {Sukhorukov}},\ and\ \bibinfo {author} {\bibfnamefont {Y.~S.}\ \bibnamefont
  {Kivshar}},\ }\bibfield  {title} {\bibinfo {title} {Scattering of linear and
  nonlinear waves in a waveguide array with a {$\mathcal{PT}$}-symmetric
  defect},\ }\href {https://doi.org/10.1103/PhysRevA.84.013833} {\bibfield
  {journal} {\bibinfo  {journal} {Phys. Rev. A}\ }\textbf {\bibinfo {volume}
  {84}},\ \bibinfo {pages} {013833} (\bibinfo {year} {2011})}\BibitemShut
  {NoStop}%
\bibitem [{\citenamefont {Zheng}()}]{zheng2026}%
  \BibitemOpen
  \bibfield  {author} {\bibinfo {author} {\bibfnamefont {C.}~\bibnamefont
  {Zheng}},\ }\href@noop {} {\bibinfo {title} {Size-dependent dynamical
  instability of periodic {$\mathcal {PT}$}-symmetric scattering systems}},\
  \Eprint {https://arxiv.org/abs/2605.10657} {arXiv:2605.10657} \BibitemShut
  {NoStop}%
\bibitem [{\citenamefont {Ladr{\'o}n De~Guevara}\ and\ \citenamefont
  {Orellana}(2006)}]{ladrondeguevara2006}%
  \BibitemOpen
  \bibfield  {author} {\bibinfo {author} {\bibfnamefont {M.~L.}\ \bibnamefont
  {Ladr{\'o}n De~Guevara}}\ and\ \bibinfo {author} {\bibfnamefont {P.~A.}\
  \bibnamefont {Orellana}},\ }\bibfield  {title} {\bibinfo {title} {Electronic
  transport through a parallel-coupled triple quantum dot molecule: {{Fano}}
  resonances and bound states in the continuum},\ }\href
  {https://doi.org/10.1103/PhysRevB.73.205303} {\bibfield  {journal} {\bibinfo
  {journal} {Phys. Rev. B}\ }\textbf {\bibinfo {volume} {73}},\ \bibinfo
  {pages} {205303} (\bibinfo {year} {2006})}\BibitemShut {NoStop}%
\bibitem [{\citenamefont {Plotnik}\ \emph {et~al.}(2011)\citenamefont
  {Plotnik}, \citenamefont {Peleg}, \citenamefont {Dreisow}, \citenamefont
  {Heinrich}, \citenamefont {Nolte}, \citenamefont {Szameit},\ and\
  \citenamefont {Segev}}]{plotnik2011}%
  \BibitemOpen
  \bibfield  {author} {\bibinfo {author} {\bibfnamefont {Y.}~\bibnamefont
  {Plotnik}}, \bibinfo {author} {\bibfnamefont {O.}~\bibnamefont {Peleg}},
  \bibinfo {author} {\bibfnamefont {F.}~\bibnamefont {Dreisow}}, \bibinfo
  {author} {\bibfnamefont {M.}~\bibnamefont {Heinrich}}, \bibinfo {author}
  {\bibfnamefont {S.}~\bibnamefont {Nolte}}, \bibinfo {author} {\bibfnamefont
  {A.}~\bibnamefont {Szameit}},\ and\ \bibinfo {author} {\bibfnamefont
  {M.}~\bibnamefont {Segev}},\ }\bibfield  {title} {\bibinfo {title}
  {Experimental observation of optical bound states in the continuum},\ }\href
  {https://doi.org/10.1103/PhysRevLett.107.183901} {\bibfield  {journal}
  {\bibinfo  {journal} {Phys. Rev. Lett.}\ }\textbf {\bibinfo {volume} {107}},\
  \bibinfo {pages} {183901} (\bibinfo {year} {2011})}\BibitemShut {NoStop}%
\bibitem [{\citenamefont {Hsu}\ \emph {et~al.}(2016)\citenamefont {Hsu},
  \citenamefont {Zhen}, \citenamefont {Stone}, \citenamefont {Joannopoulos},\
  and\ \citenamefont {Solja{\v c}i{\'c}}}]{hsu2016}%
  \BibitemOpen
  \bibfield  {author} {\bibinfo {author} {\bibfnamefont {C.~W.}\ \bibnamefont
  {Hsu}}, \bibinfo {author} {\bibfnamefont {B.}~\bibnamefont {Zhen}}, \bibinfo
  {author} {\bibfnamefont {A.~D.}\ \bibnamefont {Stone}}, \bibinfo {author}
  {\bibfnamefont {J.~D.}\ \bibnamefont {Joannopoulos}},\ and\ \bibinfo {author}
  {\bibfnamefont {M.}~\bibnamefont {Solja{\v c}i{\'c}}},\ }\bibfield  {title}
  {\bibinfo {title} {Bound states in the continuum},\ }\href
  {https://doi.org/10.1038/natrevmats.2016.48} {\bibfield  {journal} {\bibinfo
  {journal} {Nat. Rev. Mater.}\ }\textbf {\bibinfo {volume} {1}},\ \bibinfo
  {pages} {16048} (\bibinfo {year} {2016})}\BibitemShut {NoStop}%
\bibitem [{\citenamefont {Sadreev}(2021)}]{sadreev2021a}%
  \BibitemOpen
  \bibfield  {author} {\bibinfo {author} {\bibfnamefont {A.~F.}\ \bibnamefont
  {Sadreev}},\ }\bibfield  {title} {\bibinfo {title} {Interference traps waves
  in an open system: bound states in the continuum},\ }\href
  {https://doi.org/10.1088/1361-6633/abefb9} {\bibfield  {journal} {\bibinfo
  {journal} {Rep. Prog. Phys.}\ }\textbf {\bibinfo {volume} {84}},\ \bibinfo
  {pages} {055901} (\bibinfo {year} {2021})}\BibitemShut {NoStop}%
\bibitem [{\citenamefont {Siegert}(1939)}]{siegert1939}%
  \BibitemOpen
  \bibfield  {author} {\bibinfo {author} {\bibfnamefont {A.~J.~F.}\
  \bibnamefont {Siegert}},\ }\bibfield  {title} {\bibinfo {title} {On the
  derivation of the dispersion formula for nuclear reactions},\ }\href
  {https://doi.org/10.1103/PhysRev.56.750} {\bibfield  {journal} {\bibinfo
  {journal} {Phys. Rev.}\ }\textbf {\bibinfo {volume} {56}},\ \bibinfo {pages}
  {750} (\bibinfo {year} {1939})}\BibitemShut {NoStop}%
\bibitem [{\citenamefont {Hatano}(2013)}]{hatano2013}%
  \BibitemOpen
  \bibfield  {author} {\bibinfo {author} {\bibfnamefont {N.}~\bibnamefont
  {Hatano}},\ }\bibfield  {title} {\bibinfo {title} {Equivalence of the
  effective {{Hamiltonian}} approach and the {{Siegert}} boundary condition for
  resonant states},\ }\href {https://doi.org/10.1002/prop.201200064} {\bibfield
   {journal} {\bibinfo  {journal} {Fortschr. Phys.}\ }\textbf {\bibinfo
  {volume} {61}},\ \bibinfo {pages} {238} (\bibinfo {year} {2013})}\BibitemShut
  {NoStop}%
\bibitem [{\citenamefont {Sasada}\ \emph {et~al.}(2011)\citenamefont {Sasada},
  \citenamefont {Hatano},\ and\ \citenamefont {Ordonez}}]{sasada2011}%
  \BibitemOpen
  \bibfield  {author} {\bibinfo {author} {\bibfnamefont {K.}~\bibnamefont
  {Sasada}}, \bibinfo {author} {\bibfnamefont {N.}~\bibnamefont {Hatano}},\
  and\ \bibinfo {author} {\bibfnamefont {G.}~\bibnamefont {Ordonez}},\
  }\bibfield  {title} {\bibinfo {title} {Resonant spectrum analysis of the
  conductance of an open quantum system and three types of {{Fano}}
  parameter},\ }\href {https://doi.org/10.1143/JPSJ.80.104707} {\bibfield
  {journal} {\bibinfo  {journal} {J. Phys. Soc. Jpn.}\ }\textbf {\bibinfo
  {volume} {80}},\ \bibinfo {pages} {104707} (\bibinfo {year}
  {2011})}\BibitemShut {NoStop}%
\bibitem [{\citenamefont {Hatano}\ and\ \citenamefont
  {Ordonez}(2014)}]{hatano2014}%
  \BibitemOpen
  \bibfield  {author} {\bibinfo {author} {\bibfnamefont {N.}~\bibnamefont
  {Hatano}}\ and\ \bibinfo {author} {\bibfnamefont {G.}~\bibnamefont
  {Ordonez}},\ }\bibfield  {title} {\bibinfo {title} {Time-reversal symmetric
  resolution of unity without background integrals in open quantum systems},\
  }\href {https://doi.org/10.1063/1.4904200} {\bibfield  {journal} {\bibinfo
  {journal} {J. Math. Phys.}\ }\textbf {\bibinfo {volume} {55}},\ \bibinfo
  {pages} {122106} (\bibinfo {year} {2014})}\BibitemShut {NoStop}%
\bibitem [{\citenamefont {Krasnok}\ \emph {et~al.}(2019)\citenamefont
  {Krasnok}, \citenamefont {Baranov}, \citenamefont {Li}, \citenamefont {Miri},
  \citenamefont {Monticone},\ and\ \citenamefont {Al{\'u}}}]{krasnok2019}%
  \BibitemOpen
  \bibfield  {author} {\bibinfo {author} {\bibfnamefont {A.}~\bibnamefont
  {Krasnok}}, \bibinfo {author} {\bibfnamefont {D.}~\bibnamefont {Baranov}},
  \bibinfo {author} {\bibfnamefont {H.}~\bibnamefont {Li}}, \bibinfo {author}
  {\bibfnamefont {M.-A.}\ \bibnamefont {Miri}}, \bibinfo {author}
  {\bibfnamefont {F.}~\bibnamefont {Monticone}},\ and\ \bibinfo {author}
  {\bibfnamefont {A.}~\bibnamefont {Al{\'u}}},\ }\bibfield  {title} {\bibinfo
  {title} {Anomalies in light scattering},\ }\href
  {https://doi.org/10.1364/AOP.11.000892} {\bibfield  {journal} {\bibinfo
  {journal} {Adv. Opt. Photonics}\ }\textbf {\bibinfo {volume} {11}},\ \bibinfo
  {pages} {892} (\bibinfo {year} {2019})}\BibitemShut {NoStop}%
\bibitem [{\citenamefont {Huba{\v c}}\ and\ \citenamefont
  {Wilson}(2010)}]{hubac2010}%
  \BibitemOpen
  \bibfield  {author} {\bibinfo {author} {\bibfnamefont {I.}~\bibnamefont
  {Huba{\v c}}}\ and\ \bibinfo {author} {\bibfnamefont {S.}~\bibnamefont
  {Wilson}},\ }\href@noop {} {\emph {\bibinfo {title} {Brillouin-{{Wigner
  Methods}} for {{Many-Body Systems}}}}}\ (\bibinfo  {publisher} {Springer},\
  \bibinfo {address} {New York},\ \bibinfo {year} {2010})\BibitemShut {NoStop}%
\bibitem [{\citenamefont {Guevara}\ \emph {et~al.}(2003)\citenamefont
  {Guevara}, \citenamefont {Claro},\ and\ \citenamefont
  {Orellana}}]{guevara2003}%
  \BibitemOpen
  \bibfield  {author} {\bibinfo {author} {\bibfnamefont {M.~L. L.~D.}\
  \bibnamefont {Guevara}}, \bibinfo {author} {\bibfnamefont {F.}~\bibnamefont
  {Claro}},\ and\ \bibinfo {author} {\bibfnamefont {P.~A.}\ \bibnamefont
  {Orellana}},\ }\bibfield  {title} {\bibinfo {title} {Ghost {{Fano}} resonance
  in a double quantum dot molecule attached to leads},\ }\href
  {https://doi.org/10.1103/PhysRevB.67.195335} {\bibfield  {journal} {\bibinfo
  {journal} {Phys. Rev. B}\ }\textbf {\bibinfo {volume} {67}},\ \bibinfo
  {pages} {195335} (\bibinfo {year} {2003})}\BibitemShut {NoStop}%
\bibitem [{\citenamefont {Miroshnichenko}(2009)}]{miroshnichenko2009a}%
  \BibitemOpen
  \bibfield  {author} {\bibinfo {author} {\bibfnamefont {A.~E.}\ \bibnamefont
  {Miroshnichenko}},\ }\bibfield  {title} {\bibinfo {title} {Nonlinear
  {{Fano-Feshbach}} resonances},\ }\href
  {https://doi.org/10.1103/PhysRevE.79.026611} {\bibfield  {journal} {\bibinfo
  {journal} {Phys. Rev. E}\ }\textbf {\bibinfo {volume} {79}},\ \bibinfo
  {pages} {026611} (\bibinfo {year} {2009})}\BibitemShut {NoStop}%
\bibitem [{\citenamefont {Miroshnichenko}\ \emph {et~al.}(2010)\citenamefont
  {Miroshnichenko}, \citenamefont {Flach},\ and\ \citenamefont
  {Kivshar}}]{miroshnichenko2010}%
  \BibitemOpen
  \bibfield  {author} {\bibinfo {author} {\bibfnamefont {A.~E.}\ \bibnamefont
  {Miroshnichenko}}, \bibinfo {author} {\bibfnamefont {S.}~\bibnamefont
  {Flach}},\ and\ \bibinfo {author} {\bibfnamefont {Y.~S.}\ \bibnamefont
  {Kivshar}},\ }\bibfield  {title} {\bibinfo {title} {Fano resonances in
  nanoscale structures},\ }\href {https://doi.org/10.1103/RevModPhys.82.2257}
  {\bibfield  {journal} {\bibinfo  {journal} {Rev. Mod. Phys.}\ }\textbf
  {\bibinfo {volume} {82}},\ \bibinfo {pages} {2257} (\bibinfo {year}
  {2010})}\BibitemShut {NoStop}%
\bibitem [{\citenamefont {Miroshnichenko}\ and\ \citenamefont
  {Kivshar}(2005)}]{miroshnichenko2005a}%
  \BibitemOpen
  \bibfield  {author} {\bibinfo {author} {\bibfnamefont {A.~E.}\ \bibnamefont
  {Miroshnichenko}}\ and\ \bibinfo {author} {\bibfnamefont {Y.~S.}\
  \bibnamefont {Kivshar}},\ }\bibfield  {title} {\bibinfo {title} {Engineering
  {{Fano}} resonances in discrete arrays},\ }\href
  {https://doi.org/10.1103/PhysRevE.72.056611} {\bibfield  {journal} {\bibinfo
  {journal} {Phys. Rev. E}\ }\textbf {\bibinfo {volume} {72}},\ \bibinfo
  {pages} {056611} (\bibinfo {year} {2005})}\BibitemShut {NoStop}%
\bibitem [{\citenamefont {Weimann}\ \emph {et~al.}(2017)\citenamefont
  {Weimann}, \citenamefont {Kremer}, \citenamefont {Plotnik}, \citenamefont
  {Lumer}, \citenamefont {Nolte}, \citenamefont {Makris}, \citenamefont
  {Segev}, \citenamefont {Rechtsman},\ and\ \citenamefont
  {Szameit}}]{weimann2017}%
  \BibitemOpen
  \bibfield  {author} {\bibinfo {author} {\bibfnamefont {S.}~\bibnamefont
  {Weimann}}, \bibinfo {author} {\bibfnamefont {M.}~\bibnamefont {Kremer}},
  \bibinfo {author} {\bibfnamefont {Y.}~\bibnamefont {Plotnik}}, \bibinfo
  {author} {\bibfnamefont {Y.}~\bibnamefont {Lumer}}, \bibinfo {author}
  {\bibfnamefont {S.}~\bibnamefont {Nolte}}, \bibinfo {author} {\bibfnamefont
  {K.~G.}\ \bibnamefont {Makris}}, \bibinfo {author} {\bibfnamefont
  {M.}~\bibnamefont {Segev}}, \bibinfo {author} {\bibfnamefont {M.~C.}\
  \bibnamefont {Rechtsman}},\ and\ \bibinfo {author} {\bibfnamefont
  {A.}~\bibnamefont {Szameit}},\ }\bibfield  {title} {\bibinfo {title}
  {Topologically protected bound states in photonic parity--time-symmetric
  crystals},\ }\href {https://doi.org/10.1038/nmat4811} {\bibfield  {journal}
  {\bibinfo  {journal} {Nat. Mater.}\ }\textbf {\bibinfo {volume} {16}},\
  \bibinfo {pages} {433} (\bibinfo {year} {2017})}\BibitemShut {NoStop}%
\end{thebibliography}%

\end{document}